\documentclass{natureprintstyle}

\usepackage{amsfonts}
\usepackage{amssymb}
\usepackage{amsmath}

\usepackage{float}
\usepackage{placeins}
\usepackage{graphicx}

\usepackage[export]{adjustbox}

\usepackage[super,comma,sort&compress]{natbib}

\usepackage{hyperref}
\usepackage{verbatim}

\newcommand{\errUD}[2]{\ensuremath{^{+#1}_{-#2}}}
\newcommand{\XMM}{{\it XMM}}
\newcommand{\xmm}{\textit{XMM-Newton}}
\newcommand{\swift}{\textit{Swift}}
\newcommand{\nustar}{\textit{NuSTAR}}
\newcommand{\xrism}{\textit{XRISM}}
\newcommand{\xtend}{\textit{Xtend}}
\newcommand{\resolve}{\textit{Resolve}}

\title{Highly ionized clumps shooting out from a quasar at relativistic speeds}

\author{XRISM collaboration$^{*}$}

\begin{document}

\maketitle
\footnotetext[1]{
A list of participants and their affiliations appears at the end of the paper.
}

\begin{abstract}
Evidence indicates that supermassive black holes exist at the centers of virtually all galaxies. Their mass correlates with the galactic bulge mass \cite{Merritt2001}, suggesting a co-evolution with their host galaxies~\cite{KormendyHo2013}, most likely through powerful outflows/winds~\cite{Fabian2012}. 
X-ray observations have revealed the presence of highly ionized winds outflowing at sub-relativistic speeds from the accretion disks around supermassive black holes~\cite{Tombesi2010, Gofford2013}.   
However, the limited spectral resolution of current X-ray instruments has left the physical structure of the winds poorly understood, as well as their location, both of which are essential for estimating the wind kinetic power~\cite{Laha2021, Gallo2023}.
Here, the first \xrism\ observation of the luminous quasar, PDS\,456, is reported.  The high-resolution spectrometer \resolve\ onboard \xrism\ enabled the discovery of five discrete velocity components outflowing at 20--30\% of the speed of light.
This demonstrates that the wind structure is highly inhomogeneous, which is inferred to consist of up to a million clumps of size {\boldmath $(0.2\textrm{--}1.2)\times 10^{15}{\rm ~cm}$} located {\boldmath $(2\textrm{--}4)\times 10^{16}{\rm ~cm}$} away from the X-ray source. 
The mass outflow rate is estimated to be {\boldmath $60\textrm{--}170$} solar masses per year.
Subsequently, the wind kinetic power is more than {\boldmath $10^{47}{\rm \,erg\,s^{-1}}$}, exceeding the Eddington luminosity limit.
{\color{black}
This is more than enough to significantly impact the galaxy.
}
\end{abstract}


The X-Ray Imaging and Spectroscopy Mission (\xrism)~\cite{XRISM2020} is a JAXA-led international satellite in collaboration with NASA and ESA, launched on 2023 September 7 (JST) from Tanegashima, Japan.
The payload includes the X-ray calorimeter, \resolve, which has an unprecedentedly high spectral resolution of {\color{black} $\Delta E\lesssim 6$~eV (Full-Width at Half Maximum; FWHM) for most pixels} in the X-ray bandpass of 2--12~keV, providing excellent sensitivity to narrow X-ray spectral features.
\xrism\ makes it possible to resolve spectral signatures 
of relativistic winds from supermassive black holes,
enabling for the first time the detailed study of smooth/clumpy structures~\cite{Mizumoto2021}, launching radii~\cite{Yamada2024}, covering fraction~\cite{Nardini2015}, and mass and kinetic-energy outflow rates.

\xrism\ observed the nearby ($z=0.184$~\cite{Torres1997}\footnote[2]{Recent work~\cite{Bischetti2019} reported slightly different value of $z=0.185$, but the difference is too small to affect our analysis.}) 
radio-quiet quasar PDS\,456 for about 6\,days, from 2024 March 11 to 17, for a total net exposure time of 250\,ks amongst XRISM's as part of its Performance Verification program.
The black hole mass of PDS~456 is estimated to be $M_{\rm BH}\simeq 5\times 10^8M_\odot$ ($M_\odot$ is the solar mass)~\cite{Amorim2023}. Its high luminosity indicates it is accreting at the Eddington limit or even exceeding it.
PDS\,456 can be regarded as the local counterpart of accreting supermassive black holes (SMBHs) in the most active phase of galaxy-SMBH co-evolution, typically around $z\sim 2$~\cite{Heckman2014}.
Previous X-ray observations of PDS~456 highlighted the importance of this source for studying co-evolution. It persistently shows
iron K-shell absorption lines blue-shifted with velocities of $\simeq 0.3c$ ($c$ is the speed of light), a definitive sign of powerful winds~\cite{Reeves2003, Behar2010, Nardini2015, Boissay-Malaquin2019}.
In addition to X-ray observations, outflow features on galactic scales are also reported in other wavebands~\cite{Bischetti2019, Travascio2024}.
Therefore, PDS\,456 is the ideal candidate for investigating the physical nature of powerful winds from SMBHs.

\begin{figure}
    \centering
    \includegraphics[width=\hsize]{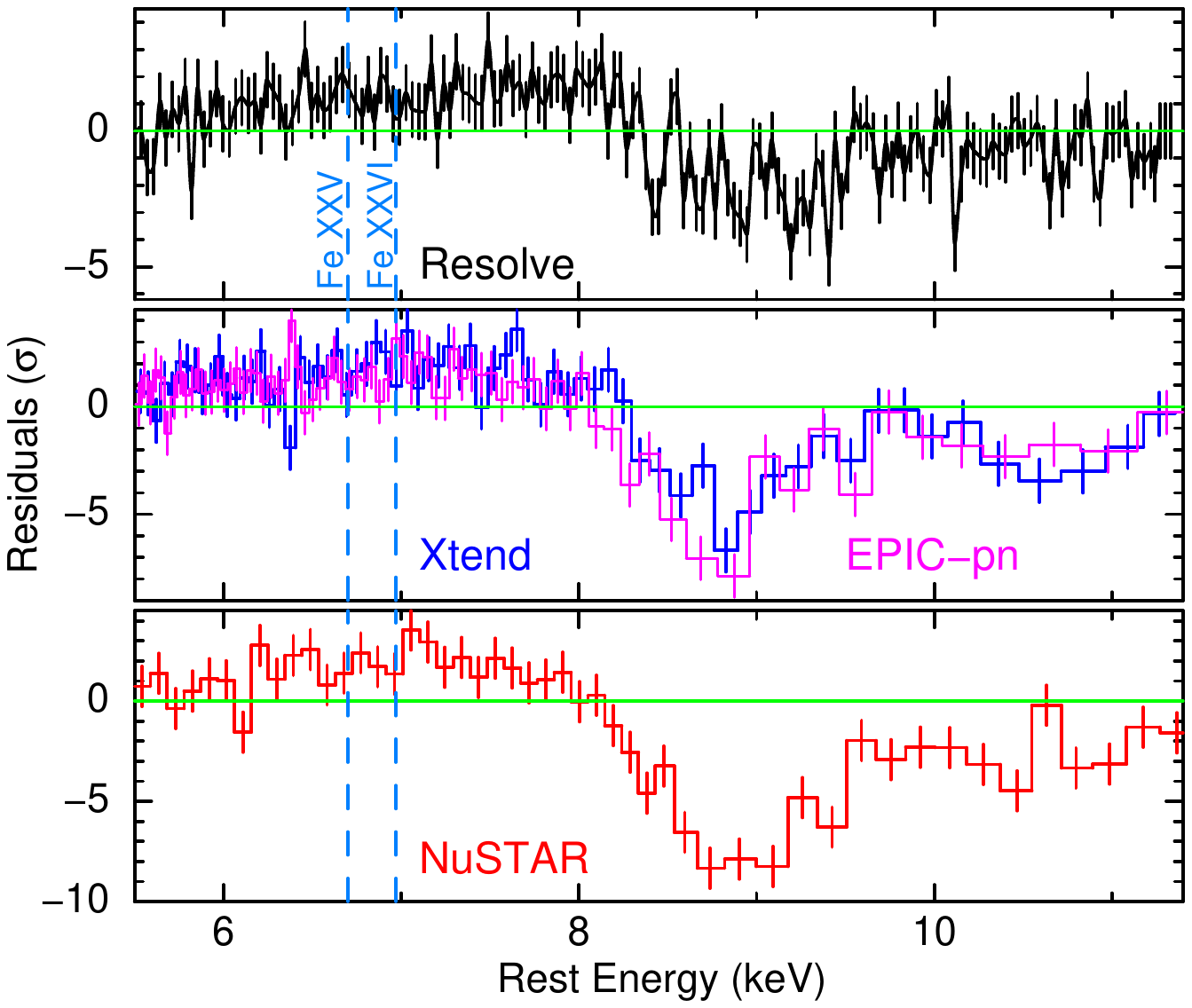}
    \caption{{\bf\label{fig:transspec} Simultaneous \xrism/\xmm/\nustar\ iron-K line profiles of PDS\,456.}
    The line profiles are shown as residuals against {a power-law continuum absorbed by Galactic and low ionization partial covering absorbers without including the Fe K emission or absorption line features.}
    Strong absorption troughs are observed between 8.4--9.4\,keV in the quasar rest frame. The expected energies of the He and H-like iron lines are at 6.70\,keV and 6.97\,keV (dashed vertical lines). This implies a range of blue-shifted velocities between $0.2c-0.3c$ in PDS\,456. The continuum model was determined by combining \xtend\ onboard \xrism\ and \nustar, which provide the broadband spectra ranging from 0.4~keV to 40~keV.
    A broad emission feature can also be discerned below 8\,keV.
    Note that the resolution at 7\,keV of \resolve\ is about 5\,eV (FWHM) corresponding to $\sim 200{\rm ~km~s^{-1}}$.
    Error bars are $1\sigma$ and energies are in the quasar's rest frame.
    {For plotting purposes, the {\it Resolve} spectra are re-binned by a factor of three, compared to a binsize of 10\,eV per bin.}
    }
\end{figure}

Figure~\ref{fig:transspec} shows the observed emission and absorption features around the iron K-shell energies obtained by \xrism, \xmm, and \nustar.
The high-resolution \resolve\ spectrum in the top panel shows clear narrow absorption lines at the rest-frame energy of 8.4--9.4~keV.
These line energies correspond to 0.22--0.33$c$ if these lines are interpreted as blue-shifted Fe~\textsc{xxv} He$\alpha$ (He-like iron), and 0.18--0.29$c$ if interpreted as Fe~\textsc{xxvi} Ly$\alpha$ (H-like iron).
These multiple absorption lines discovered by \resolve\ had up to now been observed as a single broad absorption line, because of the limited energy resolution of previous instruments.
This is demonstrated by the middle and bottom panels, which present the same plots of the simultaneous observations by the traditional charge-coupled devices \xtend\ onboard \xrism\ and \xmm, and the hard X-ray observatory \nustar.

\begin{figure}
    \centering
    \includegraphics[width=\hsize]{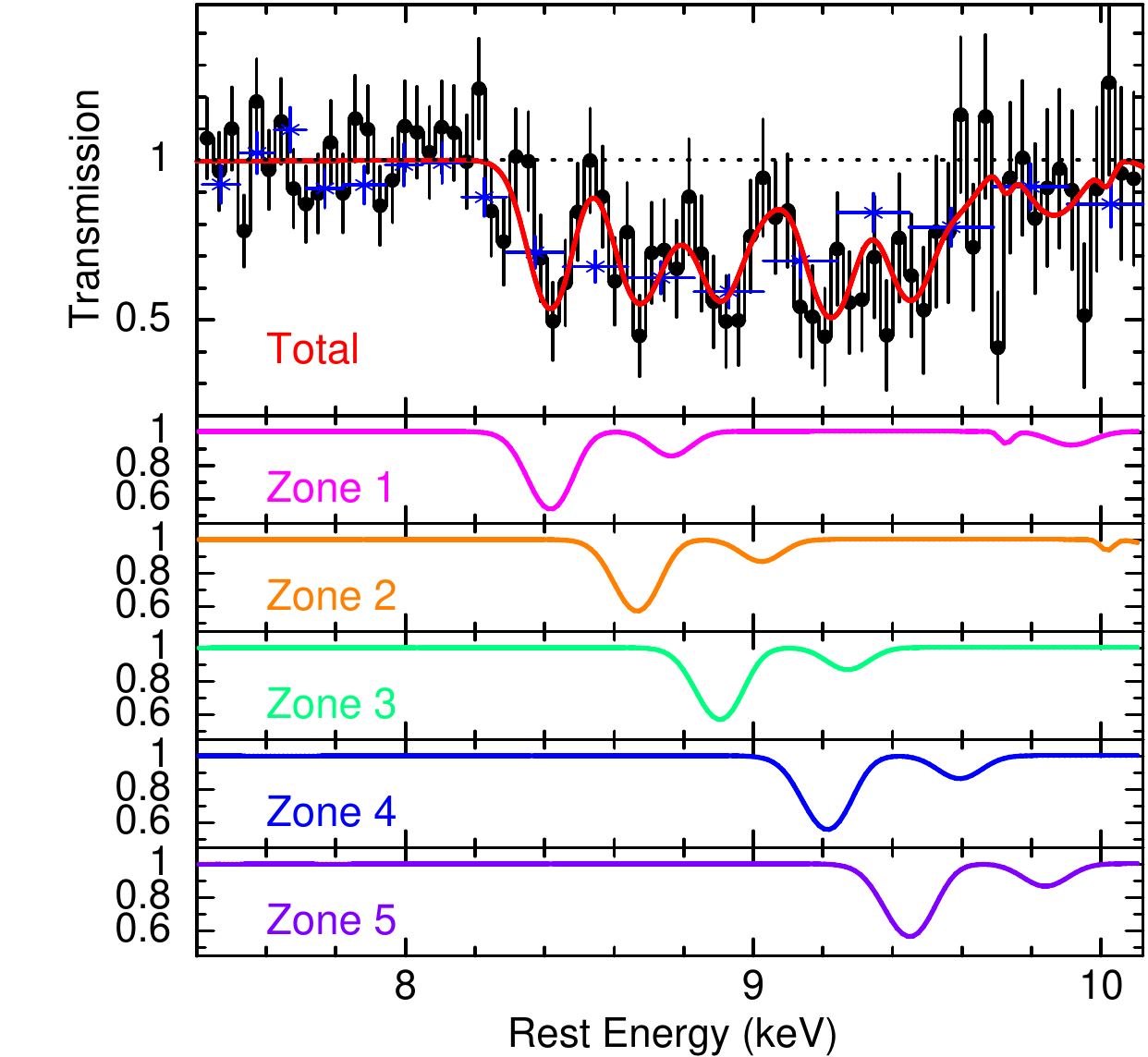}
    \caption{{\bf\label{fig:abs} Absorption models for the \xrism\ spectrum.}
    The observed spectra (\resolve\ black data points, \xtend, blue) are presented in the form of the transmission through the absorber, by dividing the data by a model consisting of an absorbed power-law continuum and the broad emission line.
    The solid red line in the upper panel shows the total transmission through all 5 absorption components combined. The lower five panels show the contributions of each component, in which the dominant opacity arises from He-like iron, with a weaker contribution from H-like iron.
    }
\end{figure}

\begin{figure}
    \centering
    \includegraphics[width=\hsize]{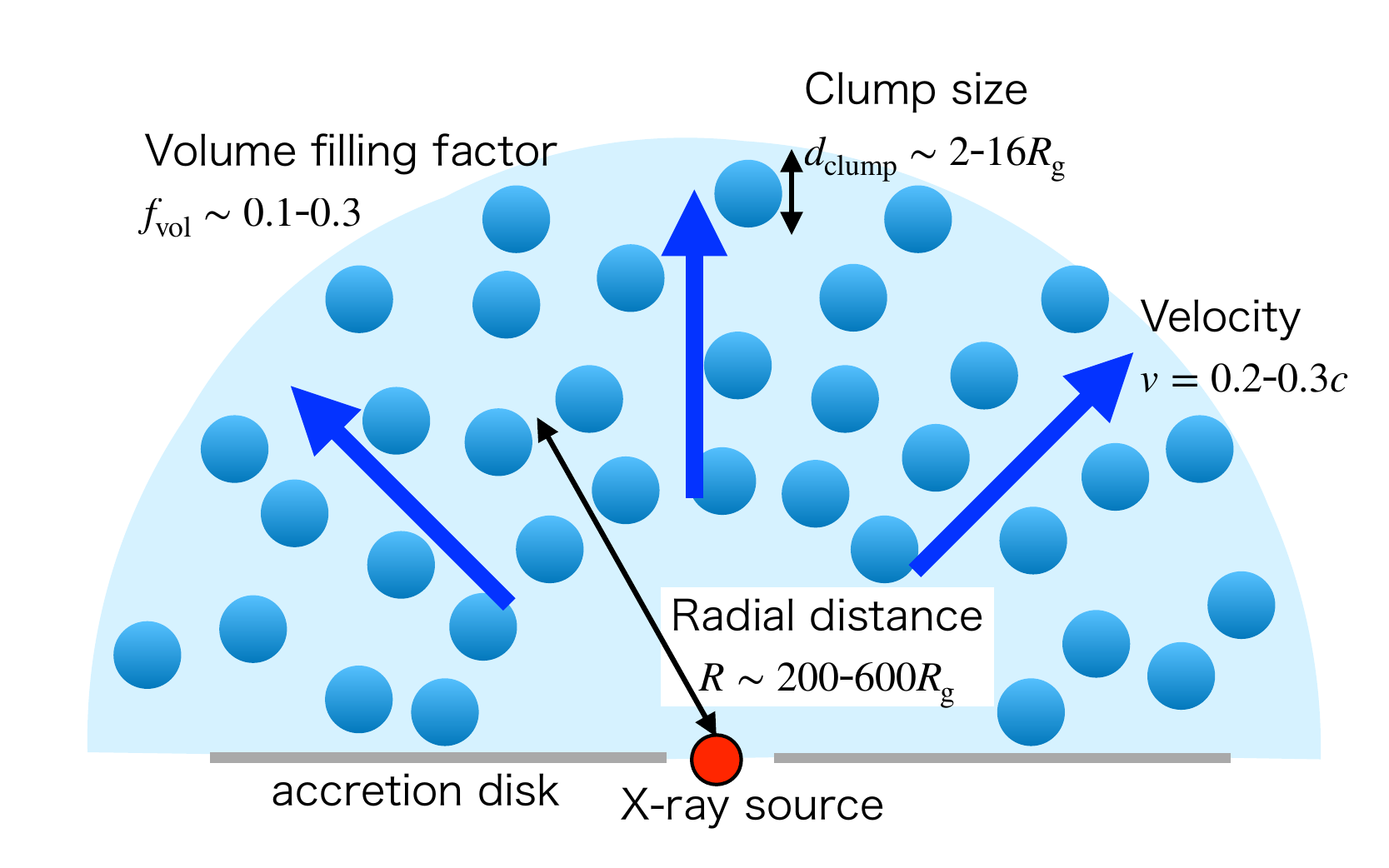}
    \caption{{\bf\label{fig:cartoon} Cartoon demonstrating the structure of the clumpy wind in PDS\,456.}
    {The blue spheres and light blue shade represent the dense clumps embedded in a less dense outflowing medium, as the simplest approximation of the inhomogeneous outflow, which in reality could look more like the figures of
    Takeuchi~et~al.~
    \cite{Takeuchi2013}}
    The \xrism\ observation suggests the outflow of PDS\,456 consists of up to a million clumps with a size of $\sim 2\textrm{--}16R_{\rm g}$, which are outflowing at $v=0.2c - 0.3c$ at a radial distance of $200\textrm{--}600R_{\rm g}$.
    Despite a volume filling factor of about 0.1--0.3, the wind fully covers the X-ray source due to the numerous clumps.
    }    
\end{figure}

We fitted the 0.4--40 keV spectra obtained by \resolve, \xtend, and simultaneous \nustar\ observation with multiple photo-ionization models with different velocities; see Methods for details. 
As shown in Fig.~\ref{fig:abs}, we invoked five absorption velocity zones, each are required at more than 99.99\% confidence {according to the Akaike Information Criteria\cite{Akaike1974, Tan2012}.}
Although the exact outflow velocities have some systematic uncertainty due to the coupling with the charge states of iron (i.e., whether they are interpreted as Fe~\textsc{xxv} or Fe~\textsc{xxvi}), the energy separation of the whole profile of about 1\,keV is too wide to be accounted for by a single velocity with multiple charge states {\color{black} because the energy separation between Fe~\textsc{i} and Fe~\textsc{xxvi} is only 0.6 keV}.
According to the modeling, the observed column densities and outflow velocities of the absorbing clumps are in the ranges of $N_{\rm H,obs}=(4.5$--$7.6)\times 10^{22}{\rm ~cm^{-2}}$ and $v_{\rm out}=0.226c -0.333c$, respectively.
Considering special relativistic effects upon the optical depths of the absorbers~\cite{Luminari2020}, the intrinsic column density ranges are slightly higher, at $N_{\rm H}=(0.8$--$1.4)\times 10^{23}{\rm ~cm^{-2}}$ for each absorber. 
To simplify the modeling, the ionization parameter and velocity dispersion were assumed to be common amongst all the absorption components.
Here, the ionization parameter $\xi$ is defined as $\xi=L_{\rm ion}/(nR^2)$, where $L_{\rm ion}$ is the 1--1000\,Rydberg ionizing luminosity, $R$ is the radial distance of the clump from the X-ray source, and $n$ is the {\color{black} electron} number density.
The best-fit values are $\log(\xi/{\rm erg~s^{-1}~cm})=4.90\pm0.14$ and $\Delta v=1900^{+600}_{-400}{\rm ~km~s^{-1}}$ (1$\sigma$ Gaussian width), respectively.
The presence of many velocity components suggests that the wind is not a smooth homogeneous flow, but rather highly inhomogeneous and clumpy as depicted in Fig.~\ref{fig:cartoon}.

\begin{figure}
    \centering
    \includegraphics[width=\hsize]{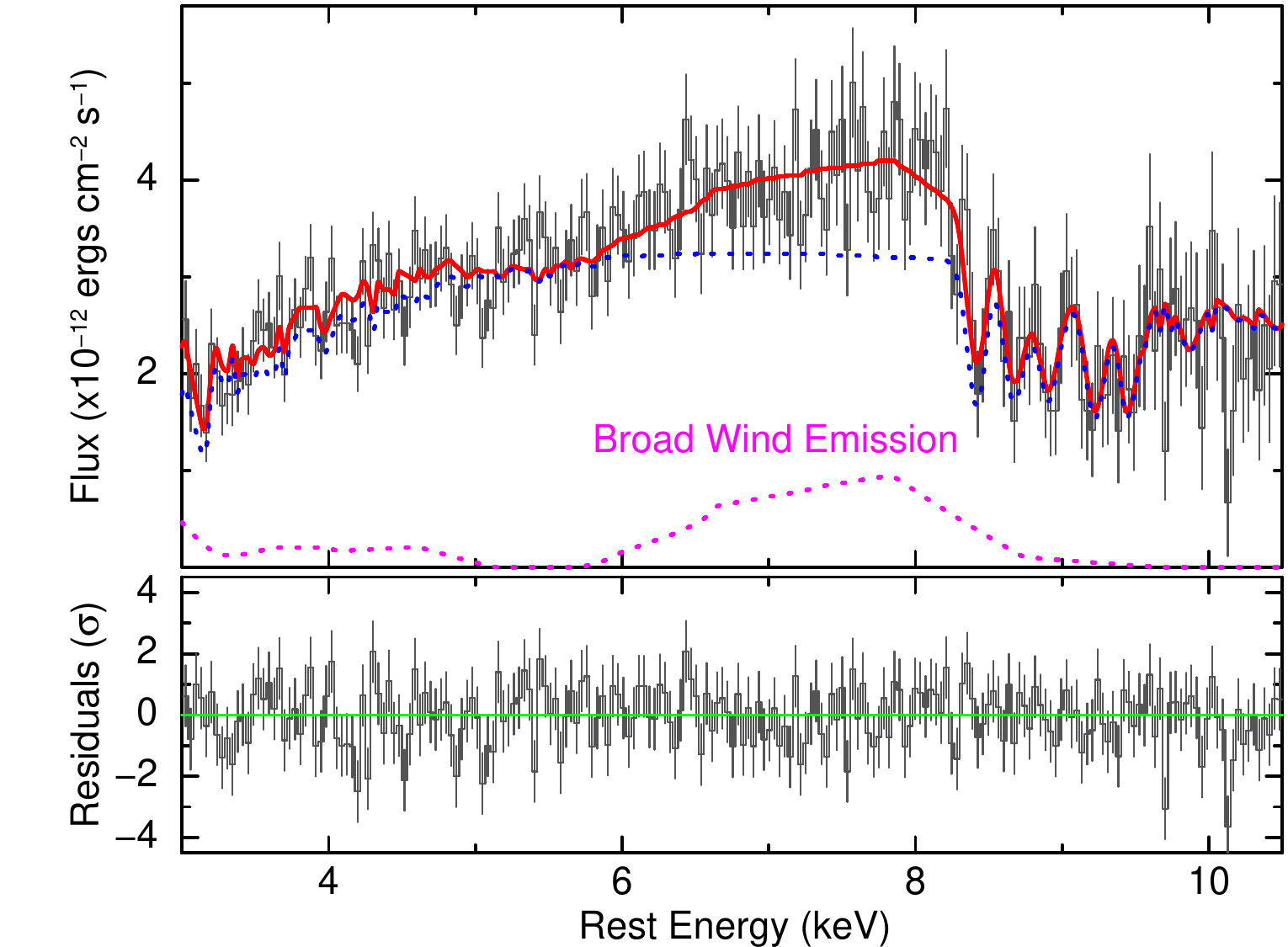}
    \caption{{\bf\label{fig:emit} Spectrum of PDS~456 in flux units observed by \xrism\ \resolve\ fitted with a self consistent emission and absorption model.}
    The emission and absorption models are calculated by the photoionized plasma model, with the same ionization parameter and column density.
    The emission model was convolved by the line profile computed from a spherical shell outflowing with velocities ranging from $0.2\textrm{--}0.3c$ (see Methods for details).
    The intensity of the emission component indicates the wind fully covers the X-ray source.
    }    
\end{figure}

PDS\,456 also features a significant broad emission line centered at $\sim7$~keV and broadened to $\sigma\sim1$~keV, corresponding to a velocity FWHM of $\sim 0.3c$, which is similar to the range of absorption velocities.
Such emission features were reported in previous studies~\cite{Nardini2015, Luminari2018, Hagino2015}.
Thus, the whole Fe-K region is naturally interpreted as a P-Cygni-like profile,
where broad emission originates from the entire wide-angle outflow.  
To self-consistently model the emission spectrum of the wind, we computed the velocity profile of a wide-angle relativistic wind with velocities of $0.2c$--$0.3c$
(see Methods for details). 
The spectra were then fitted with this emission model, assuming the ionization parameters and column densities of the emitting and absorbing plasma to be the same. Fig.~\ref{fig:emit} shows
the spectra are well accounted for by this model, demonstrating the validity of the interpretation that the emission originates from the wind. The emission model shows that the wind covering factor, compared to {$2\pi$ steradian solid angle, is $f_{\rm cov}=1.9\pm0.7$.
The value is slightly larger than unity and indicates that the wind fully covers the X-ray source (see Methods for more detail).}

The physical properties of the clumps can now be estimated, giving the best constraints on kinetic power of the wind and hence its potential impact on the host galaxy. 
The clump diameter ($d_{\rm clump}$), column density ($N_{\rm H}$) 
and number density ($n$) are related by 
$N_{\rm H} = n\, d_{\rm clump}$, while also $n=L_{\rm ion}/\xi R^2$.
From the optical to X-ray Spectral Energy Distribution of PDS\,456 (see Methods), $L_{\rm ion}=(1.6\pm0.5)\times10^{46}$\,erg\,s$^{-1}$, 
while from the spectral modeling $\log(\xi/{\rm erg\,s^{-1}\,cm})=5$ and 
$N_{\rm H}=10^{23}$\,cm$^{-2}$ per clump. This then yields $R\simeq 460R_{\rm g}\times \left(d_{\rm clump}/10R_{\rm g}\right)^{1/2}$, 
where $R_{\rm g}=GM_{\rm BH}/c^{2}\simeq 7.5\times 10^{13}{\rm ~cm}$ is the gravitational radius and $G$ is the gravitational constant.

The typical clump size and location can 
be estimated from the effect of X-ray continuum variability on the outflow (see Methods for details).
We observed a strong continuum flare during the \xrism\ observation.
Lower limits are obtained from the observational fact that the broad Fe-K emission and absorption do not vary during the \xrism\ observation, despite a factor of four change in the X-ray continuum flux.
Since the total duration of the \xrism\ observation is about 500~ks, and from the light crossing argument, the location of the clumps is estimated to be $R\gtrsim 200R_{\rm g}$ with a corresponding clump size of $d_{\rm clump}\gtrsim 2R_{\rm g}$.
On the other hand, the geometric clump size should be similar to that of the primary X-ray source, the corona, in order to absorb at least 50\% of the continuum flux between 8.4--9.4\,keV. 
According to the light-crossing argument, the doubling timescale of the flare of $\sim 40$~ks gives a maximal coronal size, setting upper limits of $d_{\rm clump}\lesssim 16R_{\rm g}$ and thus $R\lesssim 600R_{\rm g}$.
Therefore, as depicted in Fig.~\ref{fig:cartoon}, the clumps have a size of $d_{\rm clump}\sim 2\textrm{--}16R_{\rm g}\sim (1.5\textrm{--}12)\times10^{14}{\rm ~cm}$, and are located at a radial distance of $R\sim 200\textrm{--}600R_{\rm g}\sim (1.5\textrm{--}4.5)\times 10^{16}{\rm ~cm}$ from the X-ray source near the black hole.
This wind location is more accurate than those of previous studies~\cite{Gofford2014,Nardini2015}, thanks to the discovered clumpiness and long exposure time. 

The high-resolution spectra of \resolve\  allowed us to observe the number of velocity zones corresponding to the number of clumps along our line of sight, providing important information on the volume filling factor of the clumps and hence the wind mass outflow rate.
The average number of clumps along the line of sight, or ``multiplicity $M$'', is defined geometrically as the ratio $ M=N \pi (d_{\rm clump}/2)^2/(4\pi R^2)$, where $N$ is the total number of clumps in the outflow. 
The volume filling factor is $f_{\rm vol}=N(d_{\rm clump}/2)^3/R^3$. 
Combining these expressions gives $f_{\rm vol}=4M(d_{\rm clump}/2)/R\simeq 0.1\textrm{--}0.3$, for an observed value of $M=5$, corresponding to the 5 
clumps along the line of sight.
This estimation leads to the presence of a large total number of clumps in the outflow 
$N\simeq 9\times 10^{4}\textrm{--}7\times10^{5}$.

The mass outflow rate is {$\dot{M}_{\rm w}= 4\pi f_{\rm cov} f_{\rm vol} nR^2 \mu m_{p} v_{\rm out}$}, where $m_{\rm p}$ is the proton mass, $\mu=1.4$ is the mean atomic mass per proton, and from the observable parameters, $nR^2 = L_{\rm ion}/\xi = 1.6\times 10^{41}\,{\rm cm}^{-1}$ and {$f_{\rm cov}\simeq 1$.}
From the above discussion on $f_{\rm vol}$, $\dot{M}_{\rm w}\simeq 60\textrm{--}170~M_{\odot}{\rm ~yr^{-1}}$
for an average outflow velocity of $v/c=0.28\pm0.02$. 
This value is higher than previous estimate of around $10~M_{\odot}{\rm ~yr^{-1}}$~\cite{Nardini2015}, based upon the assumption of a homogeneous wind structure, but comparable to that of the molecular outflow of $290~M_{\odot}{\rm ~yr^{-1}}$~\cite{Bischetti2019}.

The Eddington luminosity of PDS\,456 is $L_{\rm Edd}\simeq 6\times 10^{46}{\rm ~erg~s^{-1}}$, while the Eddington mass accretion rate is $\dot{M}_{\rm acc}=L_{\rm Edd}/\eta c^2 \simeq 20~M_{\odot}{\rm ~yr^{-1}}$ for an efficiency factor of $\eta=0.05$. 
Even for the most conservative value of $\dot{M}_{\rm w}$, the wind kinetic power is $L_{\rm kin}=0.5\dot{M}_{\rm w}v^2\approx10^{47}$\,erg\,s$^{-1}$. Thus, the wind in PDS\,456 exceeds the Eddington luminosity.
The bolometric luminosity of PDS\,456 $L_{\rm bol}\sim 10^{47}{\rm ~erg~s^{-1}}$~\cite{Amorim2024} is also greater than $L_{\rm Edd}$.
Thus, a super-Eddington wind may not be so surprising, but see a low-Eddington example \cite{Zak2024}.
Interestingly, theoretical works on super-Eddington accretion flows suggest the presence of clumpy winds~\cite{Takeuchi2013}.
According to these works, the clumpy structure is due to the instability inherent in the radiation driving mechanism, and indeed multi-velocity winds are predicted in simulations of winds driven by radiation pressure, where UV line opacity is important~\cite{Mizumoto2021}.
The clump size predicted by these winds is $\sim 20R_{\rm g}$ at a distance of $\sim 500R_{\rm g}$, both of which are perfectly consistent with the values estimated from our observations.
We should note that the clumpy structure cannot exclude other mechanisms such as magnetic driving~\cite{Fukumura2015}, but it poses new challenges to theoretical models that need to explain the internal structure of winds as well as their extreme outflow velocity.

The \xrism\ data conclusively detect the wind in PDS~456, resolving it into a multi-velocity clumpy outflow.
The unprecedented detail revealed in this observation allows us to determine the kinetic power to be more than $10^{47}{\rm ~erg~s^{-1}}$, comparable to the bolometric luminosity.
This is large enough to explain the galaxy versus black hole co-evolution, where $L_{\rm kin}/L_{\rm bol}\simeq 0.005$ may be sufficient~\cite{Hopkins2010}.
Moreover, it is much larger than any other types of quasar winds, such as broad absorption lines (BALs) seen in ultraviolet band~\cite{Bruni2019}, low-velocity Seyfert winds
\cite{Laha2016, Gallo2023}, and molecular outflows~\cite{Lutz2020}.
Thus, winds like that of PDS~456 have the potential for the most significant impact on the host galaxy, showing feedback from the black hole in action.


\setlength{\bibsep}{0.0pt}
{\sf\small\bibliography{mybib}}

\begin{addendum}
\item [Acknowledgments] The authors thank Junjie Mao for fruitful discussions and comments, and Yuto Mochizuki for his technical advice on \textsc{XSTAR} table models.
This work was supported by JSPS KAKENHI grant numbers JP22H00158, JP22H01268, JP22K03624, JP23H04899, JP21K13963, JP24K00638, JP24K17105, JP21K13958, JP21H01095, JP23K20850, JP24H00253, JP21K03615, JP24K00677, JP20K14491, JP23H00151, JP19K21884, JP20H01947, JP20KK0071, JP23K20239, JP24K00672, JP24K17104, JP24K17093, JP20K04009, JP21H04493, JP20H01946, JP23K13154, JP19K14762, JP20H05857, and JP2 Examples of the convolved emission line p3K03459, and NASA grant numbers 80NSSC24K1148, 80NSSC24K1774, 80NSSC20K0733, 80NSSC18K0978, 80NSSC20K0883, 80NSSC20K0737, 80NSSC24K0678, 80NSSC18K1684, and 80NNSC22K1922.
LC acknowledges support from NSF award 2205918. CD acknowledges support from STFC through grant ST/T000244/1. LG acknowledges financial support from Canadian Space Agency grant 18XARMSTMA. AT is supported in part by the Kagoshima University postdoctoral research program (KU-DREAM). SY acknowledges support by the RIKEN SPDR Program. IZ acknowledges partial support from the Alfred P. Sloan Foundation through the Sloan Research Fellowship.
FT acknowledges funding from the European Union - Next Generation EU, PRIN/MUR 2022 (2022K9N5B4). JR acknowledges support from NASA \xrism\ grant 80NSSC23K0645.
Part of this work was performed under the auspices of the U.S. Department of Energy by Lawrence Livermore National Laboratory under Contract DE-AC52-07NA27344. The material is based upon work supported by NASA under award number 80GSFC21M0002. 
This work was supported by the JSPS Core-to-Core Program, JPJSCCA20220002. The material is based on work supported by the Strategic Research Center of Saitama University.
 
 \item [Author contributions] Kouichi Hagino and Ehud Behar conceived the campaign, proposed the multi-satellite observations, and led the project. Misaki Mizumoto reduced and evaluated the \resolve\ data, and  James Reeves led the data analysis. Valentina Braito and Adam Gonzalez reduced and analyzed the data of \nustar, \swift, and \xmm. Francesco Tombesi, Alfredo Luminari, and Pierpaolo Cond\'o performed the blind line search and applied photoionized wind models to the \resolve\ data. Yerong Xu performed the grid search of the photoionization model and evaluated uncertainties with different wind models. Rozenn Boissay-Malaquin, Chris Done, Luigi Gallo, Satoshi Yamada, and Francesco Tombesi contributed to the discussions in the XRISM PDS 456 target team's regular meetings. Matteo Guainazzi and Keigo Fukumura served as internal reviewers.
 The science goals of XRISM were discussed and developed over 7 years by the XRISM Science Team, all members of which are authors of this manuscript. All the instruments were prepared by the joint efforts of the team. The manuscript was subject to an internal collaboration-wide review process. All authors reviewed and approved the final version of the manuscript.
 
 \item[Competing Interests] The authors declare that they have no competing financial interests.

 \item[Data availability] The observational data analyzed during this study are partly available at NASA’s High Energy Astrophysics Science Archive Research Center (HEASARC; https://heasarc.gsfc. nasa.gov/).

  \item[Code availability] The codes used for the data reduction are available from the HEASARC website \\(https://heasarc.gsfc.nasa.gov/docs/software/heasoft) and ESA's website (https://www.cosmos.esa.int/web/xmm-newton/sas). The spectral fitting tools are freely available online (https://heasarc.gsfc.nasa.gov/xanadu/xspec for XSPEC, https://www.sron.nl/astrophysics-spex for SPEX).
 
 \item[Correspondence] Correspondence and requests for materials should be addressed to Kouichi Hagino\\(kouichi.hagino@phys.s.u-tokyo.ac.jp), Ehud Behar (behar@physics.technion.ac.il),\\
 James Reeves (james.n.reeves456@gmail.com), or Misaki Mizumoto (mizumoto-m@fukuoka-edu.ac.jp).

 \item[XRISM collaboration]
~\\
Marc Audard$^{1}$,
Hisamitsu Awaki$^{2}$,
Ralf Ballhausen$^{3,4,5}$,
Aya Bamba$^{6}$,
\\
Ehud Behar$^{7, 27}$,
Rozenn Boissay-Malaquin$^{8,4,5}$,
Laura Brenneman$^{9}$,
\\
Gregory V.\ Brown$^{10}$,
Lia Corrales$^{11}$,
Elisa Costantini$^{12}$
% ...

\begin{affiliations}
  \item Department of Astronomy, University of Geneva, Versoix CH-1290, Switzerland.
  \item Department of Physics, Ehime University, Ehime 790-8577, Japan.
  \item Department of Astronomy, University of Maryland, College Park, MD 20742, USA.
  \item NASA / Goddard Space Flight Center, Greenbelt, MD 20771, USA.
  \item Center for Research and Exploration in Space Science and Technology, NASA / GSFC (CRESST II), Greenbelt, MD 20771, USA.
  \item Department of Physics, University of Tokyo, Tokyo 113-0033, Japan.
  \item Department of Physics, Technion, Technion City, Haifa 3200003, Israel.
  \item Center for Space Science and Technology, University of Maryland, Baltimore County (UMBC), Baltimore, MD 21250, USA.
  \item Center for Astrophysics | Harvard-Smithsonian, MA 02138, USA.
  \item Lawrence Livermore National Laboratory, CA 94550, USA.
  \item Department of Astronomy, University of Michigan, MI 48109, USA.
  \item SRON Netherlands Institute for Space Research, Leiden, The Netherlands.
  \item ESO, Karl-Schwarzschild-Strasse 2, 85748, Garching bei M\"unchen, Germany.
  \item Centre for Extragalactic Astronomy, Department of Physics, University of Durham, South Road, Durham DH1 3LE, UK.
  \item Institute of Space and Astronautical Science (ISAS), Japan Aerospace Exploration Agency (JAXA), Kanagawa 252-5210, Japan.
  \item Department of Physics, Kyoto University, Kyoto 606-8502, Japan.
  \item Department of Economics, Kumamoto Gakuen University, Kumamoto 862-8680, Japan.
  \item Department of Physics, Tokyo Metropolitan University, Tokyo 192-0397, Japan.
  \item Department of Physics, Hiroshima University, Hiroshima 739-8526, Japan.
  \item Department of Physics, Fujita Health University, Aichi 470-1192, Japan.
  \item Department of Astronomy and Physics, Saint Mary's University, Nova Scotia B3H 3C3, Canada.
  \item Cahill Center for Astronomy and Astrophysics, California Institute of Technology, Pasadena, CA 91125, USA.
  \item European Space Agency (ESA), European Space Research and Technology Centre (ESTEC), 2200 AG, Noordwijk, The Netherlands.
  \item Faculty of Engineering, University of Miyazaki, Miyazaki 889-2192, Japan.
  \item RIKEN Nishina Center, Saitama 351-0198, Japan.
  \item Leiden Observatory, University of Leiden, P.O. Box 9513, NL-2300 RA, Leiden, The Netherlands.
  \item Kavli Institute for Astrophysics and Space Research, Massachusetts Institute of Technology, MA 02139, USA.
  \item Department of Physics, Saitama University, Saitama 338-8570, Japan.
  \item Department of Physics, Rikkyo University, Tokyo 171-8501, Japan.
  \item Faculty of Physics, Tokyo University of Science, Tokyo 162-8601, Japan.
  \item Faculty of Science and Technology, Tokyo University of Science, Chiba 278-8510, Japan.
  \item Department of Electronic Information Systems, Shibaura Institute of Technology, Saitama 337-8570, Japan.
  \item Department of Earth and Space Science, Osaka University, Osaka 560-0043, Japan.
  \item Department of Physics, University of Wisconsin, WI 53706, USA.
  \item Department of Physics and Astronomy, University of Waterloo, Ontario N2L 3G1, Canada.
  \item Department of Physics, Nagoya University, Aichi 464-8602, Japan.
  \item Science Research Education Unit, University of Teacher Education Fukuoka, Fukuoka 811-4192, Japan.
  \item Hiroshima Astrophysical Science Center, Hiroshima University, Hiroshima 739-8526, Japan.
  \item Department of Data Science, Tohoku Gakuin University, Miyagi 984-8588, Japan.
  \item College of Science and Engineering, Kanto Gakuin University, Kanagawa 236-8501, Japan.
  \item European Space Agency(ESA), European Space Astronomy Centre (ESAC), E-28692 Madrid, Spain.
  \item Department of Science, Faculty of Science and Engineering, KINDAI University, Osaka 577-8502, Japan.
  \item Department of Teacher Training and School Education, Nara University of Education, Nara 630-8528, Japan.
  \item Astronomical Institute, Tohoku University, Miyagi 980-8578, Japan.
  \item Department of Physics, Nara Women's University, Nara 630-8506, Japan.
  \item School of Science and Technology, Meiji University, Kanagawa, 214-8571, Japan.
  \item Yale Center for Astronomy and Astrophysics, Yale University, CT 06520-8121, USA.
  \item Department of Physics, Konan University, Hyogo 658-8501, Japan.
  \item Graduate School of Science and Engineering, Kagoshima University, Kagoshima, 890-8580, Japan.
  \item Department of Physics, Chuo University, Tokyo 112-8551, Japan.
  \item Faculty of Education, Shizuoka University, Shizuoka 422-8529, Japan.
  \item Department of Astronomy, Kyoto University, Kyoto 606-8502, Japan.
  \item Nihon Fukushi University, Shizuoka 422-8529, Japan.
  \item Anton Pannekoek Institute, the University of Amsterdam, Postbus 942491090 GE Amsterdam, The Netherlands.
  \item RIKEN Cluster for Pioneering Research, Saitama 351-0198, Japan.
  \item Johns Hopkins University, MD 21218, USA.
  \item Department of Astronomy and Astrophysics, University of Chicago, Chicago, IL 60637, USA.
  \item INAF, Osservatorio Astronomico di Brera, Via Bianchi 46 I-23807 Merate (LC), Italy.
  \item Department of Physics, Institute for Astrophysics and Computational Sciences, The Catholic University of America, Washington, DC 20064, USA.
  \item Dipartimento di Fisica, Universit`a di Trento, Via Sommarive 14, I-38123, Trento, Italy.
  \item Physics Department, Tor Vergata University of Rome, Via della Ricerca Scientifica 1, 00133 Rome, Italy.
  \item Department of Physics and Astronomy, James Madison University, Harrisonburg, VA 22807, USA.
  \item INAF, Istituto di Astrofisica e Planetologia Spaziali, Via del Fosso del Caveliere 100, I-00133 Roma, Italy.
  \item INAF, Osservatorio Astronomico di Roma, Via Frascati 33, 00078 Monteporzio, Italy.
\end{affiliations}
\end{addendum}

\newpage
\begin{methods}
\section*{Observations}
PDS 456 was observed simultaneously with six telescopes as summarized in Table~\ref{tab:obs}.
X-Ray Imaging and Spectroscopy Mission (\xrism) observed PDS~456 from 2024 March 11 to 17 over six days with a total exposure of 258~ks.
In the observation, the \resolve\ was operated without any filter with a closed gate valve, and \xtend\ was in full-window mode.
A part of this observation was also covered with simultaneous observations with two X-ray telescopes, \nustar~\cite{Harrison2013} and \xmm~\cite{Jansen2001}.
\nustar\ has co-aligned telescopes and corresponding Focal Plane Modules A (FPMA) and B (FPMB).
They have a good hard X-ray sensitivity above 10~keV, allowing us to determine the primary X-ray spectrum without spectral distortions due to absorbing materials.
On the other hand, \xmm\ has various X-ray detectors: two Reflection Grating Spectrometers (RGS)~\cite{DenHerder2001}, a European Photon Imaging Camera (EPIC) pn CCD detector~\cite{Struder2001}, and two EPIC MOS CCD detectors~\cite{Turner2001}.
Although RGS is capable of high-resolution spectroscopy at energies below 2~keV, its data does not have enough X-ray photons due to heavy absorption in the soft X-ray band.
Thus, we used only EPIC to cross-check \xtend\ data.
In addition, it has the Optical Monitor (OM) instrument~\cite{Mason2001}, which was used to model the broadband spectral energy distribution of PDS~456 described in the following section.
The observation of \nustar\ started 2~hours earlier than \xrism\ with a shorter exposure of about 160~ks, covering nearly four days of the first half of the \xrism\ observation.
The \xmm\ observation covered less than two days with an exposure of about 80~ks in the middle of the \xrism\ observation.

PDS~456 was also monitored by two X-ray telescopes, \swift~\cite{Gehrels2004} and NICER~\cite{Gendreau2016} during the \xrism\ observation.
\swift\ performed 20 observations with a typical exposure time of 1--2~ks for the X-ray Telescope (XRT)~\cite{Burrows2005} in March 2024 as listed in Table~\ref{tab:obs}.
This monitoring was conducted in a similar way to previous campaign of PDS~456~\cite{Reeves2021}, and it covers most of the \xrism\ observation period with approximately daily sampling until March 15.
Similarly, NICER also monitored from March 10 to 16 with approximately daily sampling.
This monitoring campaign provides X-ray variability with a timescale longer than the \xrism\ observation.
In addition to X-ray telescopes, simultaneous optical spectroscopic observations were also performed with the 3.8~m Seimei telescope located in Okayama prefecture of Japan~\cite{Kurita2020}.
Observations were performed with two spectrographs: Integral Field Unit of Kyoto-Okayama Optical Low-dispersion Spectrograph (KOOLS-IFU)~\cite{Matsubayashi2019}, and Tricolor CMOS Camera and Spectrograph (TriCCS).
These observations were conducted from March 9 to 15, except for March 11 and 12 when the weather was bad.
 
\section*{Data reduction of \resolve}
The \resolve\ observations were carried out with a closed gate valve, incorporating a $\sim250$~$\mu$m thick beryllium window \cite{Midooka2021}.
This limits the bandpass to energies above $\sim1.8$~keV.
Good Time Interval (GTI) filtering was applied to exclude periods during Earth's eclipse and when the sunlit Earth limb was visible, along with passages through the South Atlantic Anomaly (SAA) and times within 4300 seconds of the start of the 50-mK cooler recycling. After event screening, the combined effective exposure time was reduced to 258~ks. For the subsequent analysis, only the Hp (High-resolution Primary) events were utilized.

Data reduction was performed with the software
versions of the pre-pipeline version JAXA ``004\_002.15Oct2023\_Build7.011''  and the 
pipeline script ``03.00.011.008'', and the internal CALDB8, which corresponds to the public XRISM CALDB ver.\ 20240815. 
Events were screened based on coincidence with events in the anti-coincidence (anti-co) detector and events on other Resolve pixels.  Screening on pixel-to-pixel coincidence is implemented primarily to remove events that occur on multiple pixels when a cosmic ray deposits energy into the frame around the pixels\cite{Kilbourne2018}; an energy threshold of 300 eV was applied to avoid including crosstalk events. The standard energy-dependent rise time cut\cite{mochizuki2024} was applied.  

Time-dependent energy-scale tracking for each pixel is essential for spectroscopy with Resolve. To measure the energy scale, a set of \(^{55}\)Fe sources on the filter wheel was used to illuminate the whole array at fiducial time intervals chosen to follow the characteristic shape of the gain changes associated with the cycles of the 50-mK cooler. The Mn K$\alpha$ line complex from \(^{55}\)Fe was fitted, for each fiducial interval and each pixel, with an eight-component model with intrinsic Lorentzian profiles\cite{Holzer1997} convolved with Gaussian smoothing representing the detector resolution. Our observations include 16 fiducial measurements to track and correct the energy scale.  Correction of the energy scale is achieved via nonlinear interpolation between gain curves measured at different heat-sink temperatures\cite{porter16} and linear interpolation between fiducial measurements. Among the 36 pixels in the Resolve detector, 35 are usable for scientific purposes, and one is a calibration pixel outside of the aperture. However, Pixel 27 is subject to random gain jumps of unknown origin lasting $\sim$1--2 hours. Because the current fiducial strategy cannot track these jumps, we have excluded data from Pixel 27. 
Gaussian modeling of the detector’s line spread {\color{black}for the Mn K$\alpha$ line} indicates an instrument resolution of $4.55\pm0.05$~eV Full-Width Half Maximum (FWHM), with an energy shift of $\lesssim0.1$~eV.  Early calibration measurements indicate that the energy scale is accurate to within approximately $\pm1$~eV from 2--8 keV\cite{porter2024,eckart2024}.

When a subset of the event grades is selected for analysis, the Response Matrix File (RMF) ordinarily needs to be scaled by the fraction of events retained.  For Resolve data, however, many events that saturate the output of the signal digitizer (produced by shallow-angle cosmic rays) get flagged as one Lp (low-resolution primary) and multiple false Ls (low-resolution secondary) events through the action of the secondary-pulse detection algorithm on a clipped pulse.  In our observations, the count rate ranged from $\sim10^{-3}$--$10^{-2}$~s$^{-1}$~pix$^{-1}$, thus almost all events should have a Hp grade.  Thus, we removed Ls events from the cleaned event file before using rslmkrmf to produce an RMF, with a parameter file of xa\_rsl\_rmfparam\_20190101v006.fits.
The following line-spread-function components were included: the Gaussian core, exponential tail to low energy, escape peaks, silicon fluorescence, and electron loss continuum (i.e., the "X" option was selected). An auxiliary response file (ARF) was generated using the xaarfgen task, assuming a point-like source at the aim point as input.

\section*{Non X-ray background of \resolve}
Evaluation of the Non-X-ray background (NXB) for \resolve\ was carried out following the methods recommended by the \xrism\ team.
The \xrism\ team provided a stacked NXB event file based on night-earth observation data collected from 2023-12-25 to 2024-04-15, and a spectral model to describe the NXB spectrum. 
This ``baseline'' model consists of a continuum component approximated by a power law and emission line components modeled by 17 narrow positive Gaussians, representing Al-K$\alpha$1/K$\alpha$2, Au-M$\alpha$1, Cr-K$\alpha$1/K$\alpha$2, Mn-K$\alpha$1/K$\alpha$2, Fe-K$\alpha$1/K$\alpha$2, Ni-K$\alpha$1/K$\alpha$2, Cu-K$\alpha$1/K$\alpha$2, Au-L$\alpha$1/L$\alpha$2, and Au-L$\beta$1/L$\beta$2. Although each line is essentially composed of multiple Lorentzians broadened by the line spread function, they are approximated as Gaussians due to the insufficient statistical quality of the background data to justify more detailed descriptions.
The sigmas of the Gaussian models for the K$\alpha$1 and K$\alpha$2 lines for each element were tied and fixed, and the line intensity ratio of K$\alpha$1 to K$\alpha$2 was set at 2:1 for each element.

Next, we derived the NXB spectrum corresponding to the observation period of PDS 456. Since the NXB flux varies with Cut-Off Rigidity (COR), it was necessary to extract the NXB file corresponding to the COR values during the observation. This calculation was performed using a \texttt{rslnxbgen} task.
This task generates a COR exposure time histogram corresponding to GTI of the events, and then calculates the weighted NXB spectrum based on the histogram. The generated NXB event file was screened using the same criteria as the \resolve\ data, excluding pixel 27 and using only the Hp grade. This process produced an NXB spectrum that can be directly compared with the source spectra, with an exposure time of 807~ks.

Finally, we constructed a model to reproduce the NXB spectrum. Although the ratio of the continuum to line components is not expected to change significantly, slight variations are possible. Additionally, since the NXB excluding the Mn emission lines originates from cosmic rays whereas the Mn lines originate from the $^{55}$Fe calibration source, their normalization might vary independently. Therefore, to fit the extracted NXB spectrum, we allowed three parameters from the ``baseline'' model to vary: the normalization of the power law, the emission line excluding Mn, and the Mn lines. In the model fitting, we used an RMF file with a unity diagonal matrix. The resultant NXB spectrum and the best-fit model is shown in Figure \ref{fig:rslnxb}, and the NXB model compared to the PDS 456 source spectrum is shown in Figure \ref{fig:rslnxb_src}.

In the figures for the \resolve\ spectrum in the main text, the NXB-subtracted spectra are shown for the sake of visibility. It should be noted that, however, the model fitting throughout this paper is performed for the non-NXB-subtracted spectrum, and the fitting model has both the model to explain the source spectrum and the NXB model determined in this section.
The model fitting was performed with 10~eV binning and with C statistics.

\begin{figure}
    \centering
    \includegraphics[width=\hsize]{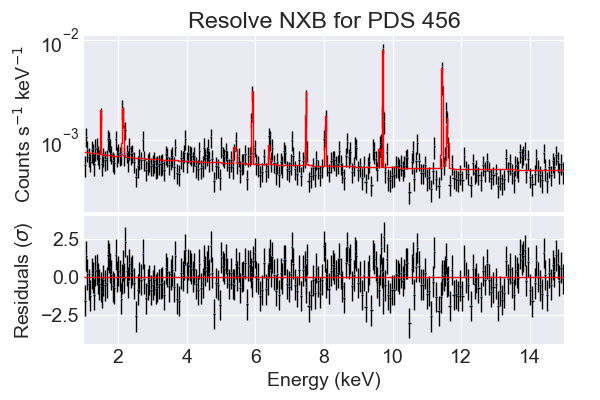}
    \caption{{\bf\label{fig:rslnxb} \resolve\ NXB spectrum and its best-fit model.} The upper panel shows the spectrum (black) and model (red), whereas the lower one shows the residuals between the data and the model.
    The spectrum is binned for the showing purpose.}    
\end{figure}
\begin{figure}
  \centering
    \includegraphics[width=\hsize]{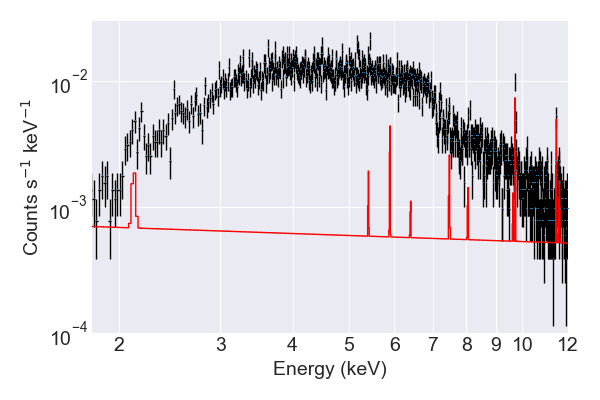}
    \caption{{\bf\label{fig:rslnxb_src} PDS 456 spectrum with the \resolve\ NXB model.} The black bins shows the PDS 456 spectrum, and the red line shows the NXB model.
    The spectrum is binned for the showing purpose.}    
\end{figure}

\section*{Data reduction of \xtend, \nustar, \xmm, and \swift}

In addition to the \resolve\ data, we reduced the observational data of \xtend, the CCD imaging spectrometer onboard \xrism, using the \xrism\ team's internal version of the \textsc{HEASoft} tools and calibration files, which will be soon made public.
We extracted the \xtend\ spectra from a circular region with a radius of $2.'5$ for the source, and a box region of $10.'3\times 9.'1$ excluding a circular region with a radius of $2.'7$ for the background.
We generated the redistribution matrix file with \textsc{xtdrmf} and the ancillary response file with \textsc{xaarfgen}.

A deep observation with the Nuclear Spectroscopic Telescope Array  (\nustar~\cite{Harrison2013})  was coordinated with the \xrism\ observation, starting two hours beforehand and ending on March 15th (see Table 1). We reduced the \nustar\ data following the standard prescriptions and using the  Data Analysis Software  (\textsc{nustardas}, ver. 2.1.2) developed for the \nustar\ mission. We used the most recent calibration files available in the Calibration Database (CALDB  version 20240520). We filtered the observation applying the standard screening criteria{~\cite{Forster2014}}, where we filtered for the passages through the SAA  setting the mode to  ``optimised"  in \textsc{nucalsaa} and the option ``tentacle = yes". 

 For each of the Focal Plane Module (FPMA and FPMB)  we extracted the source spectra using a circular region with a $50''$ radius, while for the background spectra  we adopted a region with a $76''$ radius located on the same detector.  The same regions were used to extract light curves in  different energy bands using the \textsc{nuproducts} task and adopting an orbital binsize of 5814 s.  We note that the  FPMB detector is affected by a stray light issue, which is caused by a bright source outside the FOV; however, the stray light contamination does not affect the on-axis detector.
After checking for consistency, we also combined the spectra and responses from the individual FPMA and FPMB detectors into a single spectrum, which was binned to at least 100 counts per bin. For the spectral fitting, we considered the 3--45 keV energy range, as above 45 keV the spectrum becomes background dominated.   This yields a combined  net  count rate of $0.328\pm 0.002$ counts s$^{-1}$.  The FPMA and FPMB background subtracted light curves were also combined into a single one. 

An \xmm\ observation of  PDS\,456 was also coordinated with the \xrism\ pointing. The observation started on March 13th and lasted a full orbit ($\sim 130$ ks) ending on March 15th. The  \xmm--EPIC instruments operated in small window mode and had the thin filter applied. Here, we concentrated on the EPIC-pn data, which have the highest signal-to-noise in the 2-10 keV band, although the MOS1 and MOS2 were also inspected for consistency. The raw event files were processed and cleaned using the Science Analysis Software  (SAS ver. 20.0.0~\cite{SAS}). 
 The EPIC data were first filtered for high background, which  affected only a few  ks at the end of the observation. The resulting net exposure time after correcting for dead time is $\sim 83.1$ ks. The EPIC-pn  source  spectrum  was extracted  using a
circular region  with a   radius of $30''$, while for the background we adopted two circular regions  with a radius of $30''$, respectively.  The response matrices and  the ancillary response files at the source position  were generated  using the SAS tasks \textsc{arfgen} and \textsc{rmfgen} and the latest calibration available.  The EPIC-pn net source count rate in the 0.3--10 keV band is   $0.703\pm 0.003$ counts s$^{-1}$. The pn source spectrum was   binned to have at least 50 counts  in each energy bin. 

The Optical Monitor \cite[OM][]{Mason2001} was operated in Image + Fast mode using all 6 on-board filters (\textit{UVW2}, \textit{UVM2}, \textit{UVW1}, \textit{U}, \textit{B}, and \textit{V}). We used the \textsc{omichain} task with default inputs to extract corrected images for each exposure, from which a single spectral file was created using the \textsc{om2pha} task. The canned OM response matrices are used during spectral fitting\footnote{The canned OM response matrices are available at: \href{https://www.cosmos.esa.int/web/xmm-newton/om-response-files}{https://www.cosmos.esa.int/web/xmm-newton/om-response-files}.}.

The \swift\ X-ray Telescope (XRT~\cite{Burrows2005}) data as listed in Table~1 were processed with v0.13.7 of the \textsc{xrtpipeline} to create the lightcurves and spectra. A circular source extraction region of 20$^{\prime\prime}$ was used, while for the background 
an annulus of radii 40$^{\prime\prime}$ and 130$^{\prime\prime}$ centered on the source was adopted. During the monitoring, the background subtracted XRT rates varied from $0.02$--$0.22$\,counts\,s$^{-1}$ over the 0.3--10\,keV band. Each XRT pointing was also accompanied by exposures with the \swift\ UVOT (Ultra Violet Optical Telescope~\cite{uvot}), which provided photometric measurements in the V and UVW1 bands. 

\begin{figure*}
\begin{center}
\includegraphics[width=0.45\hsize]{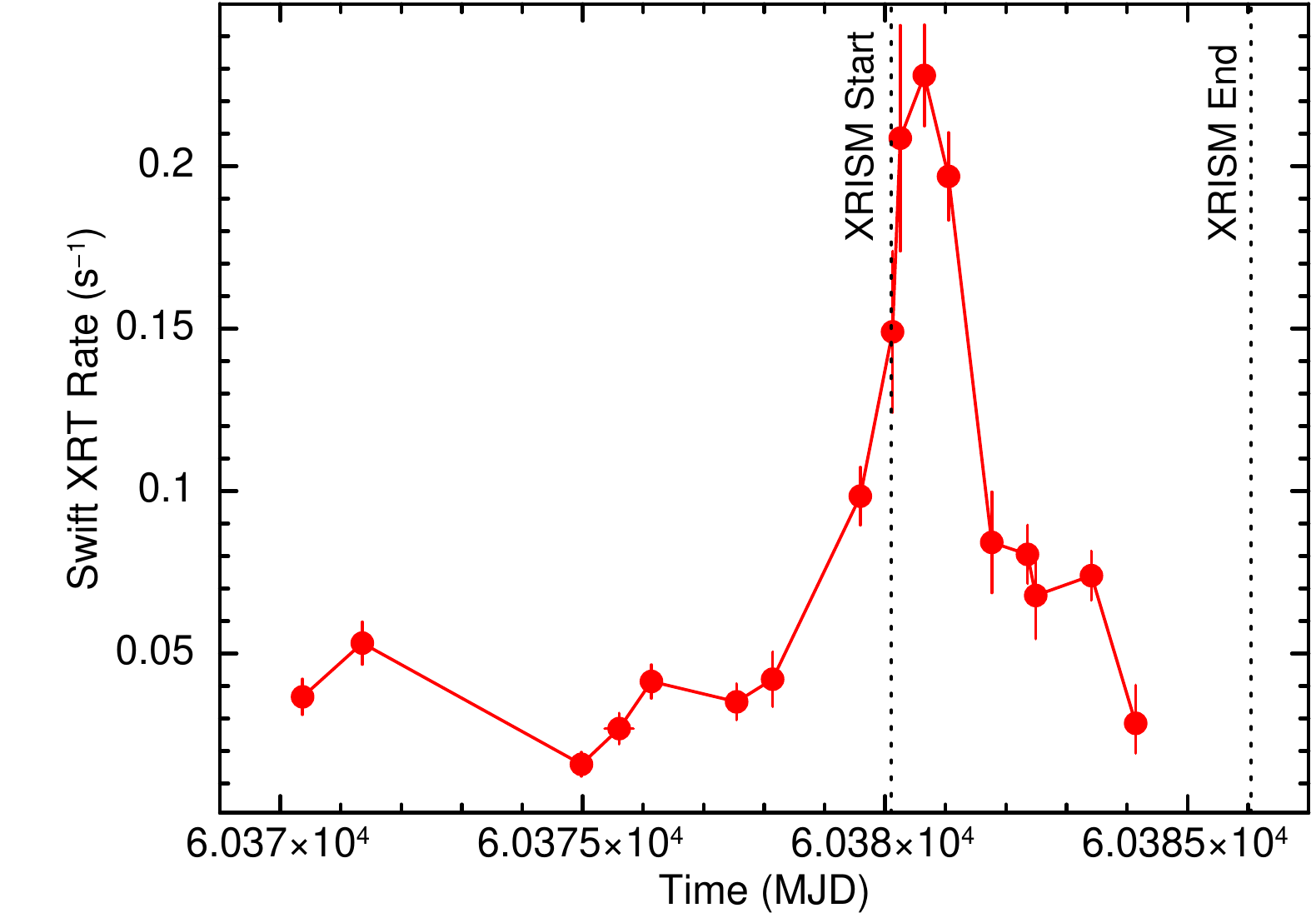}
\includegraphics[width=0.45\hsize]{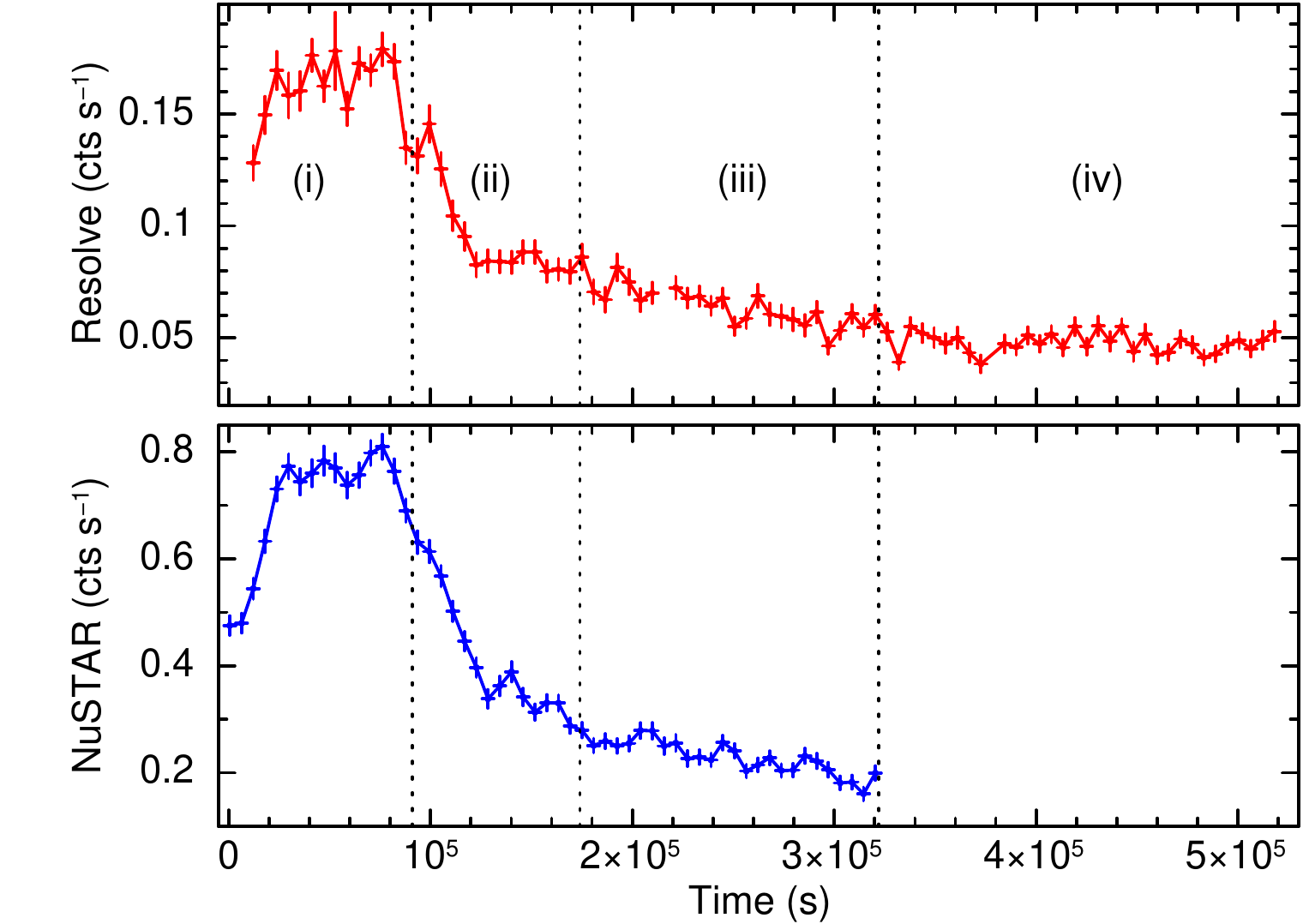}
 \end{center}
\caption{{\bf\label{fig:lightcurves} X-ray light curves of PDS\,456 from the 2024 campaign.} The left panel shows the \swift\ XRT light curve from the period up to and including the \xrism\ observation, where the intervals for the latter are marked by the vertical dotted lines. Time is plotted by Modified Julian Date (MJD). 
A large X-ray flare occurs two days prior to the start of the \xrism\ observation, where the flux increased by a factor of about $\times 5$. Prior to this, PDS\,456 was in a historically low state compared to past observations. The right panel shows the corresponding light curves taken with \xrism\ \resolve\ and \nustar\ over the 2--10\,keV and 3--40\,keV bands, where $t=0$ corresponds to the start of the \nustar\ observation. 
The peak of the \swift\ flare was caught by both observatories in interval (i). Intervals (ii) to (iv) then follow the decline in flux. The \xmm\ observation was performed in interval (iii). }
\end{figure*}

\section*{X-ray variability}
Figure~\ref{fig:lightcurves} shows the net \swift\ XRT light curve collected over the 0.3--10\,keV band, using the snapshots listed in Table~1. In the first portion of the lightcurve, the count rates from PDS\,456 were low, ranging from 
$0.02$--$0.05$\,counts\,s$^{-1}$. In comparison, PDS\,456 typically varies by an order of magnitude from $0.05$--$0.5$\,counts\,s$^{-1}$ from past \swift\ monitoring campaigns; see Reeves~et~al.~\cite{Reeves2021} for details. 
This suggests that PDS\,456 was caught in an unusually low state prior to the \xrism\ observation. Fortuitously, 2 days prior to the \xrism\ observation, the flux of PDS\,456 increased by at least a factor of $\times5$ over a 2 day period and where the maximum in the \swift\ XRT rate corresponded to the initial start of the \xrism\ and \nustar\ observations.
XRT light curves were also extracted over the soft ($0.3$--$1.5$\,keV) and hard X-ray ($1.5$--$10$\,keV) bands, although the hardness ratio between these bands show no variations before, during or after the peak of the flare. 
Note that there was no intrinsic variability of PDS\,456 seen in the \swift\ UVOT and the PDS\,456 flux variations are thus driven by the X-ray continuum.

\begin{figure*}
\begin{center}
\hspace{-1cm}
\includegraphics[width=0.45\hsize]{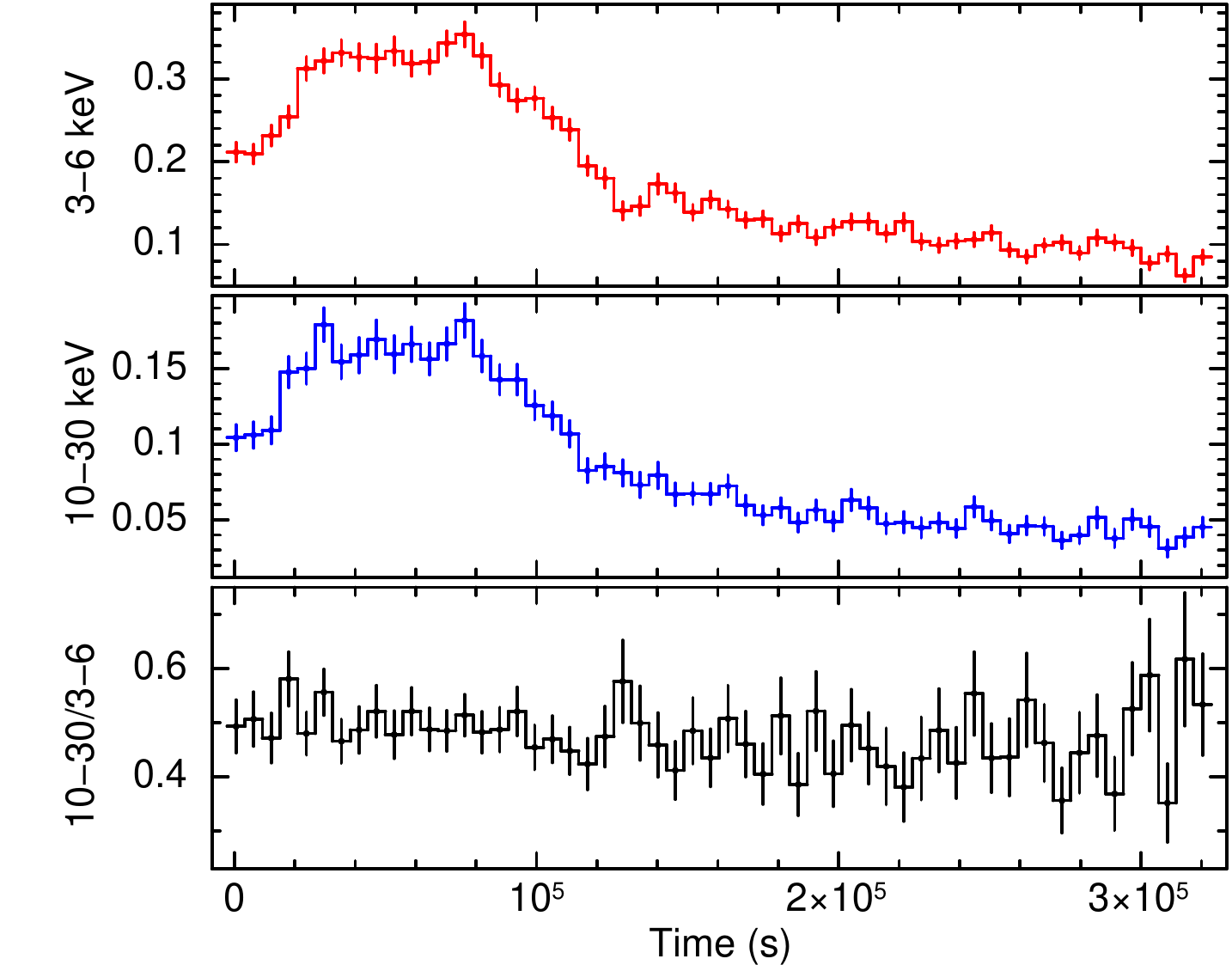}
\includegraphics[width=0.45\hsize]{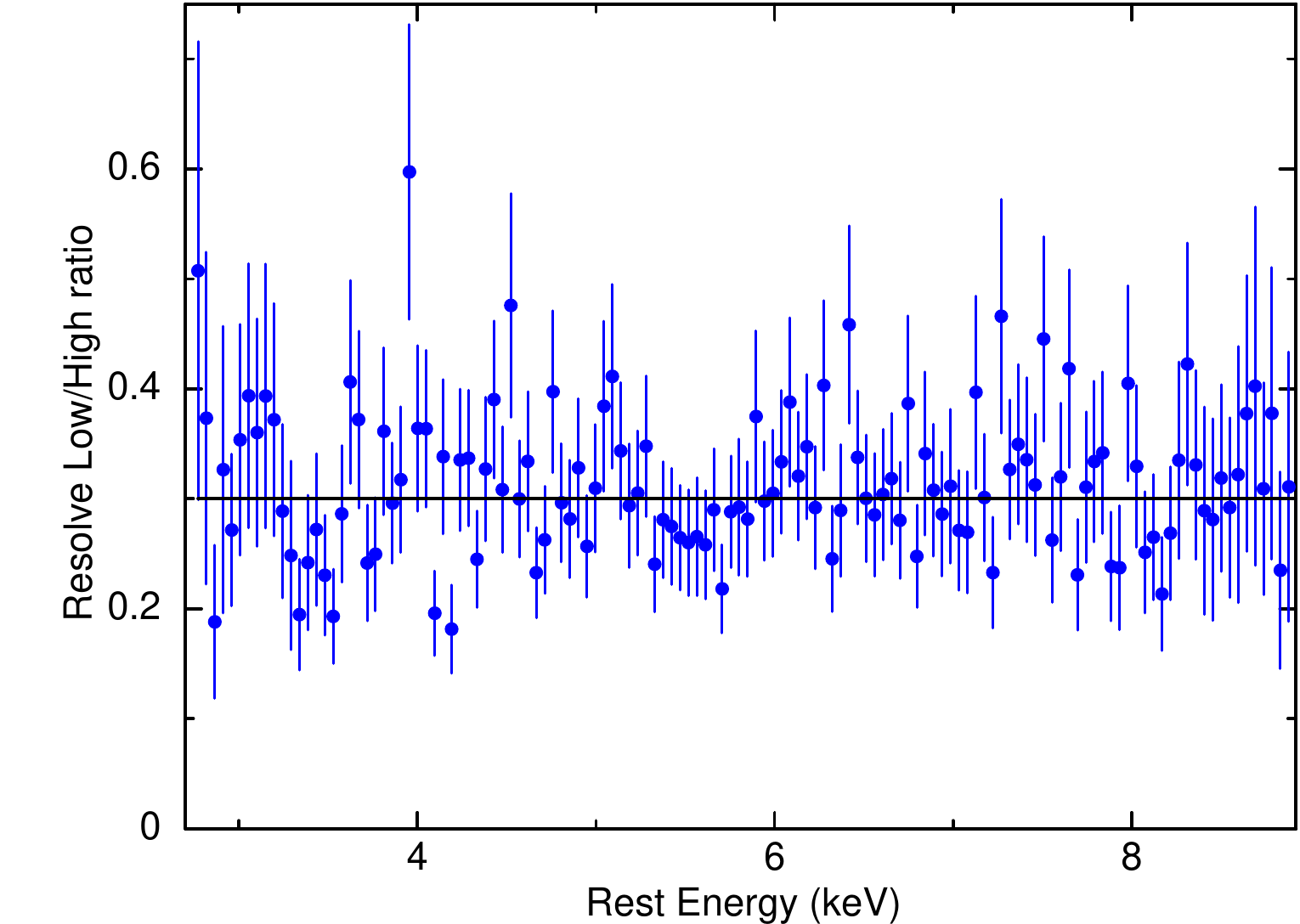}
 \end{center}
\caption{{\bf\label{fig:nustar_hardness} Spectral variability of PDS\,456.} The left panels show the \nustar\ light curves and hardness ratio.  The panels from top and middle panels show the 3--6\,keV, 10--30 keV band curves and the lower panel the hardness ratio ($HR_{\nustar}$) light curve, where  $HR_{\nustar}$ was defined as $CR_{\mathrm{10\textrm{--}30\, keV}}/CR_{\mathrm{3\textrm{--}6\, keV}}$. Note that despite the strong flux variability there is no accompanying changes in the $HR$. Indeed, the lower panel shows the lack of any spectral variability over the whole observation. The right panel shows the ratio of the low to high flux \resolve\ spectra from 3--9\,keV over the iron K band. The ratio spectrum is consistent with a constant factor of 0.3, which also implies there is no strong spectral variability of the emission or absorption over the duration of the observations.}
\end{figure*}

The light curves were compared between \xrism\ \resolve\ and \nustar, over the 2--10\,keV and 3--40\,keV bands respectively. Figure~\ref{fig:lightcurves} (right panels) shows that both observatories caught the peak of the flare and then showed a gradual decline in flux thereafter, with most of the X-ray variability occurring within the first 2 days (or 170\,ks) of the start of the observations. The light curves were also sliced into four intervals following the changes in flux; e.g. (i) X-ray flare, (ii) steep flux decline, (iii) gradual decline and (iv) X-ray quiet period. The first three intervals were coincident between \xrism\ and \nustar\ and interval (iii) had simultaneous coverage with \xmm. 
Interval (iv) had no simultaneous coverage with either \nustar\ or \xmm. During the observations, the 2--10\,keV source flux declined from an average of $9.0\times10^{-12}$\,erg\,cm$^{-2}$\,s$^{-1}$ in interval (i) down to $1.9\times10^{-12}$\,erg\,cm$^{-2}$\,s$^{-1}$  in interval (iv). 
Despite this large change in flux, no apparent hardness ratio variations were seen. Figure~\ref{fig:nustar_hardness} shows the \nustar\ light curves from 3--6 keV (top panel), 10--30 keV (middle panel) and the hardness ratio (defined as $HR_{\nustar}= CR_{\rm{10\textrm{--}30\, keV}}/CR_{\rm{3\textrm{--}6\, keV}}$; lower panel).  
The hardness ratio, reveals no variability in the spectral shape during the whole observation and that the X-ray flare is basically ``colorless".  This in turn means that the flare is driven solely by the intensity of the primary emission and not by variations in the X-ray absorber or by changes in X-ray photon index.  

Furthermore, the spectral variability was investigated by creating a ratio of the spectra from the low versus high flux intervals for the \resolve\ data. Interval (i), corresponding to the X-ray flare, was chosen for the high flux spectrum, while intervals (ii) to (iv) were extracted for the low flux part. The resulting ratio, plotted over the 3--9\,keV iron K band, is also shown on the right hand panel of Figure~\ref{fig:nustar_hardness}. The ratio spectrum shows no trend with energy nor any emission or absorption features and is consistent with a constant ratio of $\times0.3$ between the low and high flux spectra. This is also consistent with ``colorless" variability, with only the normalization of the power-law continuum varying through the observations.
 
\begin{figure}
\begin{center}
\includegraphics[width=\hsize]{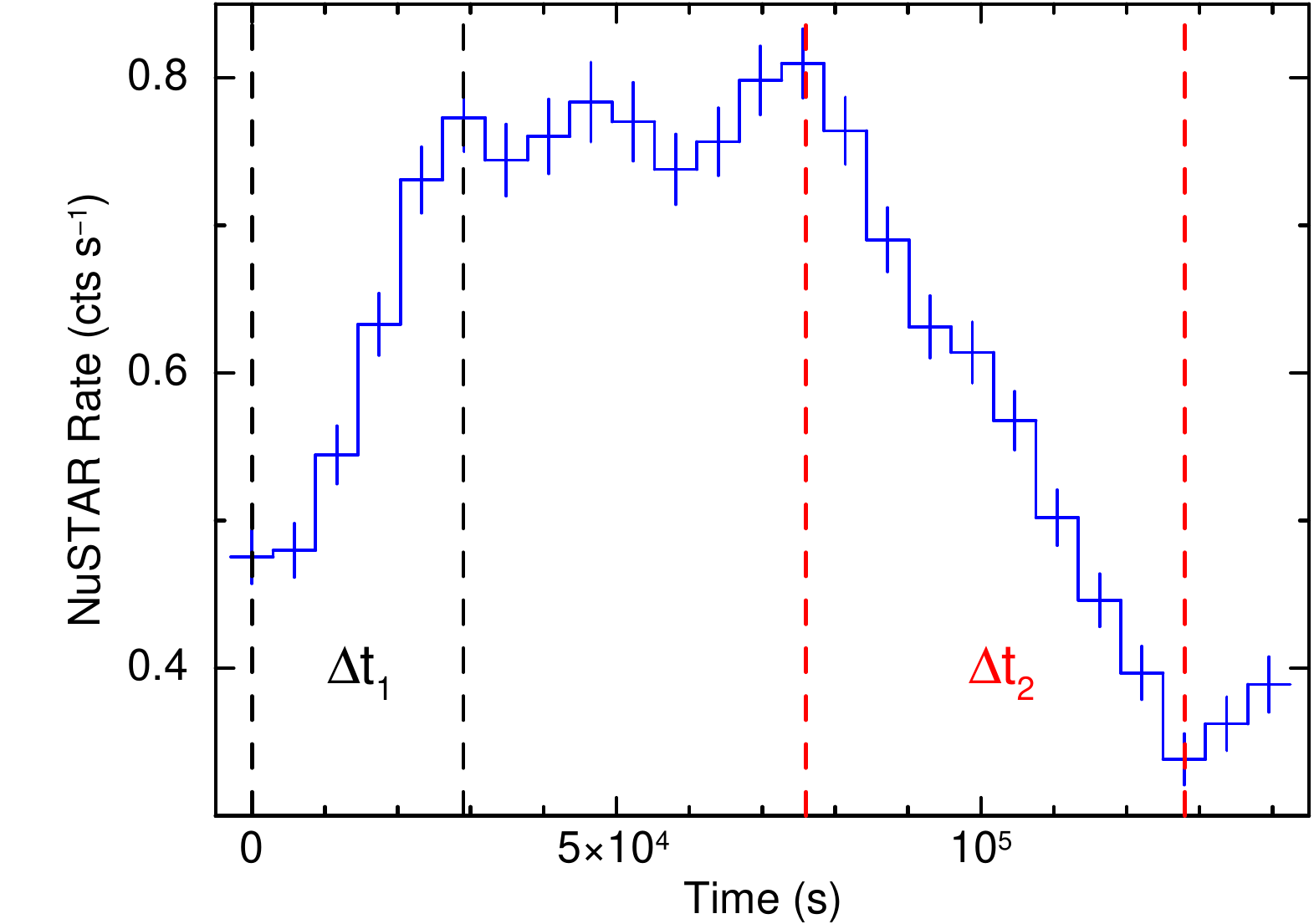}
 \end{center}
\caption{{\bf\label{fig:nustar_flare} Zoom in over the 3--40\,keV \nustar\ light curve over the first 2 days of the observation, in order to highlight the variability.} The interval $\Delta t_1$ marks the rise portion of the curve, while $\Delta t_2$ indicates the decline portion. From these the timescale for the source flux to double (or half) was estimated to be 40\,ks and implies an emission region of size $D=1.2\times10^{15}$\,cm from the light crossing argument.}
\end{figure}

The 3--40\,keV \nustar\ light curve was used to measure the rate of increase or decline of the X-ray flux and thus the time taken for the flux to double or half in value. From the light crossing considerations, this gives an estimate of the absolute size of the varying region, the X-ray corona.   A close up of the first 2 days of the \nustar\ light curve is shown in Figure~\ref{fig:nustar_flare}. The first interval ($\Delta t_1$) marks the rise portion of the curve, where the rates increase from $0.47\pm0.02$ to $0.77\pm0.02$\,counts\,s$^{-1}$ over the first 29\,ks of the observation. For a constant rate of increase, this implies a doubling time of $45\pm5$\,ks or  
$t_{\rm double}=38\pm4$\,ks in the rest frame of PDS\,456 at $z=0.184$. An equivalent calculation over the decline portion of the light curve ($\Delta t_2$) yields a halving time of $t_{\rm half}=37\pm3$\,ks. 
Thus, a characteristic variability timescale of 40\,ks corresponds to a light crossing distance of $D= c\Delta t \approx 1.2\times10^{15}$\,cm. For a black hole mass in PDS 456 of $M_{\rm BH}=5\times10^{8}\,{\rm M}_{\odot}$, then the estimated coronal size is $D\approx16 R_{\rm g}$ in gravitational units. 

\section*{The spectral energy distribution of PDS 456 and photoionization models}
The spectral energy distribution (SED) of PDS\,456 was then defined for use as an input to generate photoionized emission and absorber models for the spectral analysis. 
The ionization state is sensitive to the exact distribution of the ionizing photons above 13.6\,eV. In particular, one of the most critical parameters is the X-ray photon index, which sets the ionization balance for highly ionized Fe K lines. Also important is the slope of the UV to X-ray continuum, which determines the fraction of high energy photons capable of ionizing the wind in hard X-rays.  

\begin{figure}
\begin{center}
\includegraphics[width=\hsize]{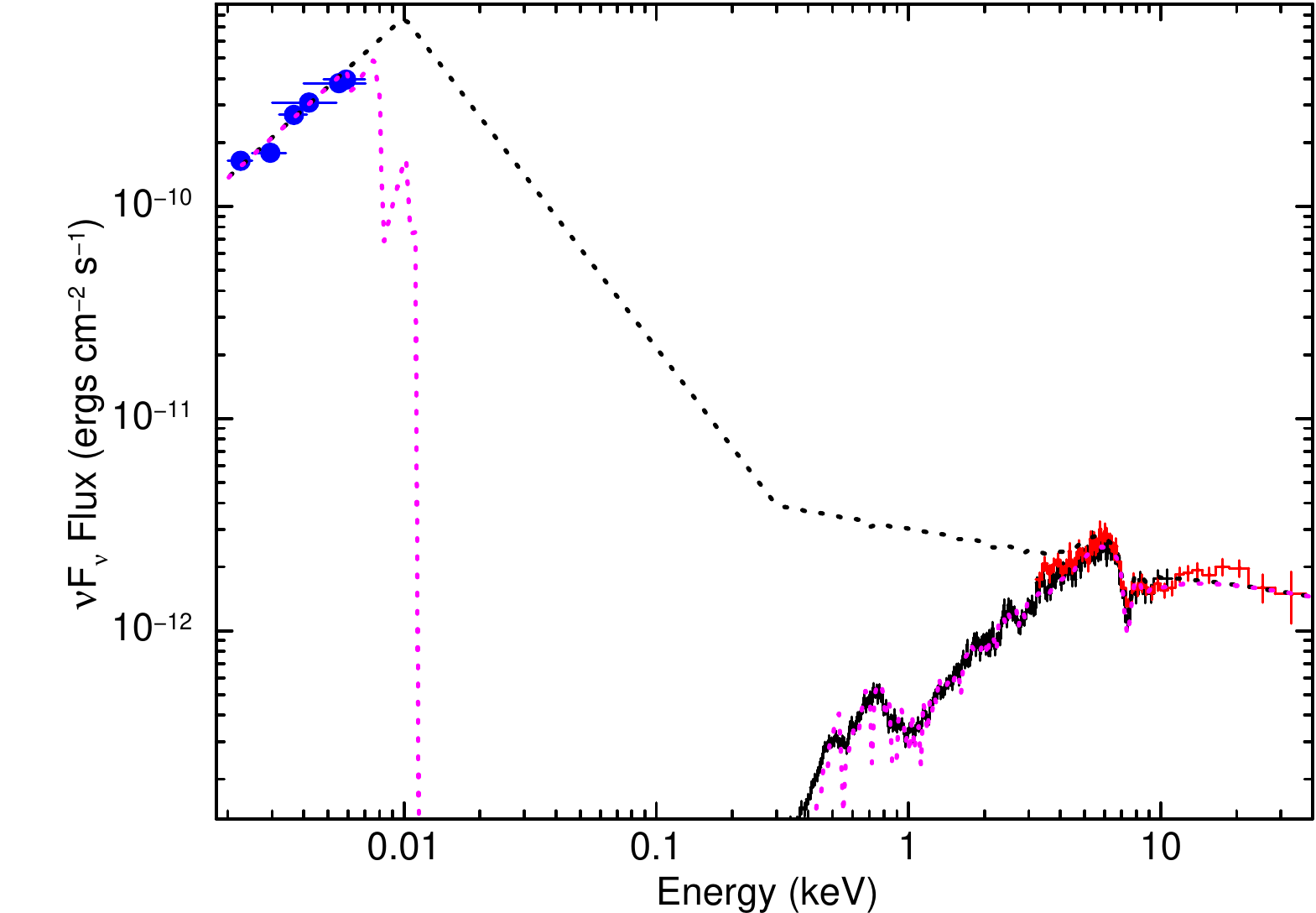}
 \end{center}
\caption{{\bf\label{fig:sed} The SED of PDS\,456 from the 2024 observations.} The black data points are from \xmm\ EPIC-pn, the red points are from the simultaneous \nustar\ interval (iii) spectrum and the OM photometry points are shown in blue. Note the latter are corrected for Galactic reddening. A simple double broken power-law function is used to describe the intrinsic continuum from the optical to hard X-rays (black dotted curve), while the SED without correction for absorption, as fitted to the data, is shown as the magenta dotted curve. 
After correction for absorption, the 1--1000\,Rydberg band luminosity of PDS\,456 is $5\times10^{46}$\,erg\,s$^{-1}$.}
\end{figure}

To define the SED, simultaneous optical to X-ray data were utilized comprising the 6 \XMM\ OM photometric bands (see data reduction above), the EPIC-pn spectrum and the \nustar\ spectrum extracted in the interval (iii), as defined above.  
The flux in each of the \xmm-OM bands was also corrected for the Galactic reddening towards PDS\,456 of $E(B-V)=0.51$. The X-ray absorption due to neutral line of sight matter in our Galaxy, of 
$N_{\rm H}=3.4\pm0.5\times10^{21}$\,cm$^{-2}$, was accounted for using the \textsc{tbabs} \textsc{xspec} model and  abundances of Wilms~et~al.~\cite{Wilms2000}. A simple parameterized form of the SED was adopted, consisting of a double broken power-law function with breaks at 10\,eV and 300\,eV. 
This makes it possible to parameterize the slope of the optical to UV continuum (up to 10\,eV), the UV to X-ray slope (10--300\,eV) and the X-ray photon index ($0.3$--$40$\,keV). In this epoch, PDS\,456 also shows pronounced intrinsic soft X-ray absorption which is also accounted for in the SED modeling (Figure~\ref{fig:sed}); i.e. the intrinsic or unabsorbed SED is derived. The details of the soft X-ray absorption are given in the spectral modeling section. 

The SED fit to interval (iii) yielded photon indices of $\Gamma_{\rm opt\textrm{-}uv}=0.91\pm0.05$ ($<10$\,eV), $\Gamma_{\rm uv\textrm{-}x}=3.50\pm0.04$ ($10$--$300$\,eV) and $\Gamma_{\rm x}=2.26\pm0.05$ ($0.3$--$40$\,keV). 
The X-ray luminosity over the 2--10\,keV band is $L_{2\textrm{--}10\,{\rm keV}}=(3.8\pm0.1)\times10^{44}$\,erg\,s$^{-1}$, while the 1--1000\,Rydberg band ionizing luminosity is $L_{\rm ion}=(4.2\pm0.8)\times10^{46}$\,erg\,s$^{-1}$. 
The data and SED model are shown in Figure~\ref{fig:sed}. 

The optical and UV band fluxes for PDS\,456 are not known to vary on timescales of days~\cite{Reeves2021}, which is also confirmed by the \swift\ UVOT monitoring performed around the \xrism\ campaign. As a result, 
the  \xmm\ OM photometric points can be used alongside the average \xrism-\xtend\ and \nustar\ data to derive the average SED during the campaign. 
The main difference compared to interval (iii) is the average X-ray luminosity, which is somewhat higher due to the initial X-ray flare, with  $L_{2\textrm{--}10\,{\rm keV}}=(6.4\pm0.2)\times10^{44}$\,erg\,s$^{-1}$. This 
also results in a flatter UV to X-ray slope of $\Gamma_{\rm uv\textrm{-}x}=3.29\pm0.04$. The other SED parameters remained the same and overall the ionizing luminosity is consistent within the uncertainties, with $L_{\rm ion}=(5.1\pm1.5)\times 10^{46}$\,erg\,s$^{-1}$.
The actual luminosity, as seen in the reference frame of the outflowing gas, will then be lower compared to this value, accounting for the relativistic de-boosting upon the continuum. 
Following equation~2 in Luminari~et~al.~\cite{Luminari2020}, for a mean velocity of $v/c=-0.28$, then the subsequent deboosting factor is $\Phi=0.32$, resulting in a de-boosted ionizing luminosity of $L_{\rm ion}=(1.6\pm0.5)\times 10^{46}$\,erg\,s$^{-1}$.
 
This SED is used as an input into the  \textsc{xstar}  photoionization code~\cite{xstar} in order to generate high spectral resolution grids of  photo-ionized emitters and absorbers, with a resolution of 2\,eV at 6\,keV. 
A gas number density of $n=10^{8}$\,cm$^{-3}$ is adopted, {though the models for Fe-K} are not sensitive to this parameter. A variety of grids were generated to cover a wide range in column density ($N_{\rm H}= 10^{22} \textrm{--} 10^{24}$\,cm$^{-2}$), ionization parameter ($\log(\xi /{\rm erg\,cm \,s^{-1})}=2$--$7$), and turbulence velocity ($v_{\rm turb}$ from 500 km s$^{-1}$ to 9000 km s$^{-1}$) in a similar way to Mochizuki~et~al.~\cite{Mochizuki2023}. This made it possible to simultaneously model both the 
high ionization iron K band emission and absorption, as well as the lower ionization soft X-ray gas. To predict the emission from a wide angle wind over a wide range in velocities, the photoionized emission spectra were then convolved with the emission line profiles as computed below.

{
We note that using the same SED for all the absorbing zones is reasonable, because the effect of shadowing on the Fe-K region due to the other absorbers is considered to be small. 
Highly-ionized absorbers
do not change the ionizing continuum significantly. On the other hand, a low-ionization absorber can change the ionizing continuum, but not much above the Fe-K edge at $\sim$7~keV, which determines the Fe-K ionization states.
}

\section*{Emission line profile}
\begin{figure}
    \centering
    \includegraphics[width=\hsize]{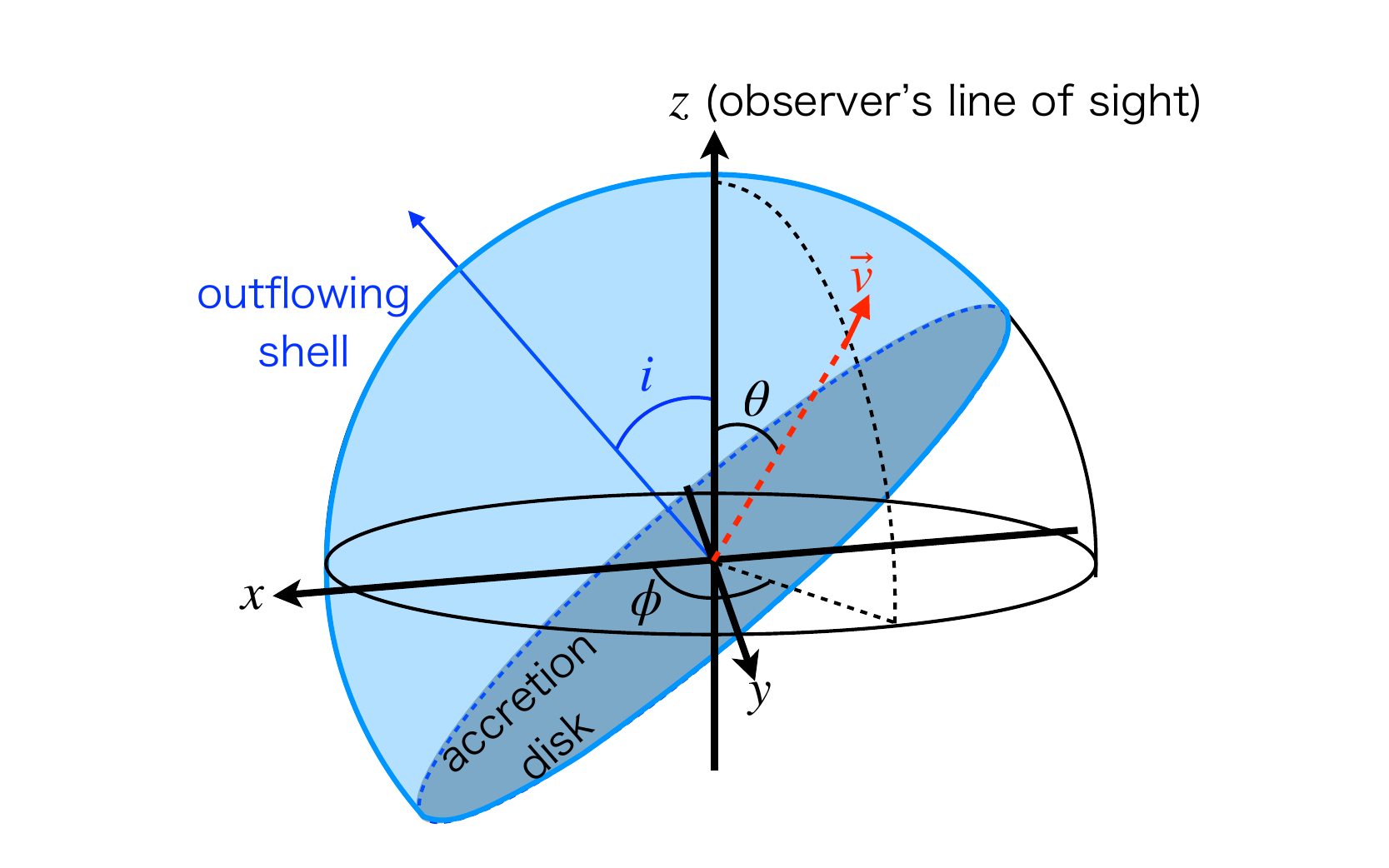}
    \caption{
    {\bf\label{fig:emission_geometry} Geometry and coordinates assumed for the emission line profile.}
    A hemispherical shell radially outflowing is assumed for the calculation of the emission line profile. The blue hemisphere represents the outflowing shell above the accretion disk, which is inclined with an angle $i$.
    }    
\end{figure}

The emission model was constructed by convolving the \textsc{xstar} emission model with a Doppler-broadened line profile under a simple velocity structure.
The \textsc{xstar} emission model used the same ionizing SED as the \textsc{xstar} absorption tables and the same grids of ionization parameters and column densities.
{\color{black}
In addition to the wind emission, relativistic reflections from the inner accretion disk could contribute to this emission feature, but previous studies have shown that most of this emission is due to wind and it is unlikely that the disk reflections contribute significantly~\cite{Nardini2015}.
}

The line profile for the convolution was calculated by assuming a spherical thin shell outflowing with a constant velocity in a radial direction {as shown in Fig.~\ref{fig:emission_geometry}}.
{
Despite the volume filling factor of
0.1--0.3, the X-ray source is fully covered due to the large number of clumps; on average five clumps along each line of sight.
Thus, the spherical shell geometry is a good approximation.
The clumpy structure might affect the intensity of the scattered/reflected light, like the case of the clumpy torus~\cite{Tanimoto2019}.
However, this second-order effect is beyond the scope of 
this paper.
}
We only consider emission from the outflow in the near side of the accretion disk, because we expect the disk would block photons from the other side.
The observer's line of sight is $\theta=0$ in the spherical coordinate system.
The disk is inclined by $i$ around the axis of $\phi=\pi/2$ (i.e., the $y$-axis in Cartesian coordinates, see Fig.~\ref{fig:emission_geometry}).

To calculate the line profile, we considered special relativistic effects, including Doppler shift $E=\delta E_0$ and Doppler beaming $L(E)=\delta^3 L_0(E_0)$, where $E$ is the photon energy, $L(E)$ is the luminosity density at energy $E$, and subscript $0$ indicates those in the rest-frame.
The Doppler factor $\delta$ is expressed as $\delta=\sqrt{1-\beta^2}/(1-\beta\cos\theta)$.
By considering these effects, the line profile $L(E)\equiv dL/dE$ for the input monochromatic emission line at an energy $E_{0}$ with a luminosity $L_{0}$ can be calculated as 
\begin{align}
L(E)&=\frac{dL}{d\cos\theta}\frac{d\cos\theta}{dE}=\delta^3L_0w_\phi(\theta,i)\frac{d\cos\theta}{dE}\nonumber\\
&=L_0w_\phi(\theta,i)\frac{\sqrt{1-\beta^2}}{\beta E_0^2}E.
\end{align}
Here, $w_\phi(\theta,i)$ is the integral over the azimuthal angle $\phi$ in the outflowing hemispherical shell and is solved analytically as
\begin{align}
w_\phi(\theta,i)&\equiv \frac{1}{2\pi}\int_{\cos\phi<\cot\theta\cot i} d\phi\nonumber\\
&=
\begin{cases}
1 & (0<\tan\theta\leq \cot i) \\
0 & (-\cot i<\tan\theta<0) \\
1-\frac{\cos^{-1}\left(\cot\theta\cot i\right)}{\pi} & (\cot i<\tan\theta,\tan\theta<-\cot i)
\end{cases}
.
\end{align}
If written as photon flux (i.e., in units of ${\rm ph~s^{-1}~cm^{-2}~keV^{-1}}$), the profile for the input monochromatic emission line with a photon flux of $N_{0}$ becomes
\begin{align}
N(E)=\frac{L(E)}{E}=N_0w_\phi(\theta,i)\frac{\sqrt{1-\beta^2}}{\beta E_0}. \label{eq:conv}
\end{align}

The resultant line profile is shown in the left panel of Fig.~\ref{fig:conv}. We assumed an inclination angle of $i=15^{\circ}$ based on a recent estimation by the GRAVITY collaboration~\cite{Amorim2024}, and velocities ranging from $0.2c$ to $0.3c$ to approximately match those of the absorption lines.
{The present emission model is thus consistent with the result of the GRAVITY collaboration~\cite{Amorim2024}, while relativistic disk reflection models require contradicting high inclination angles~\cite{Reeves2021, Nardini2015, Chiang2017}.}
Although redward shifted photons below 7~keV
are blocked by the disk in this nearly face-on configuration with $i = 15^{\circ}$,
{as shown in the right panel of Fig.~\ref{fig:conv}, red-shifted photons come from receding parts of the wind on the observed side of the disk.}
This example of a profile for a 7.0~keV line is used to convolve all of the emission features calculated by \textsc{xstar}.

\begin{figure*}
    \centering
    \includegraphics[width=0.45\hsize]{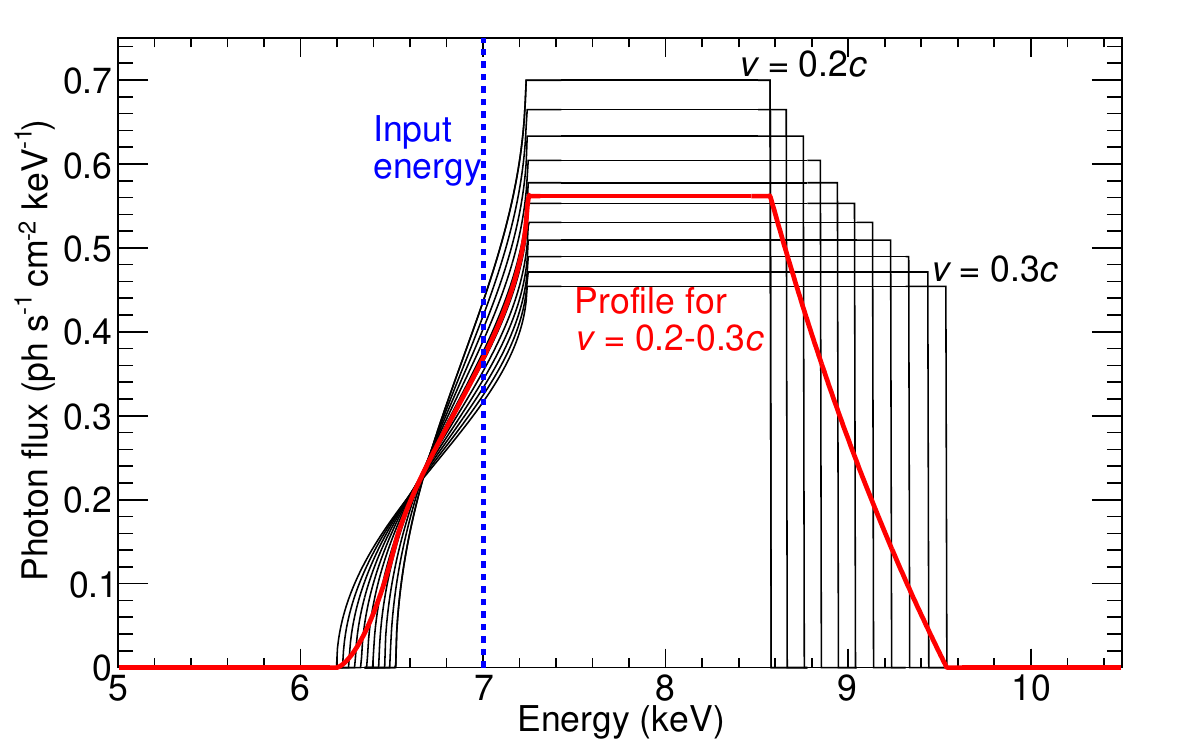}
    \includegraphics[width=0.45\hsize]{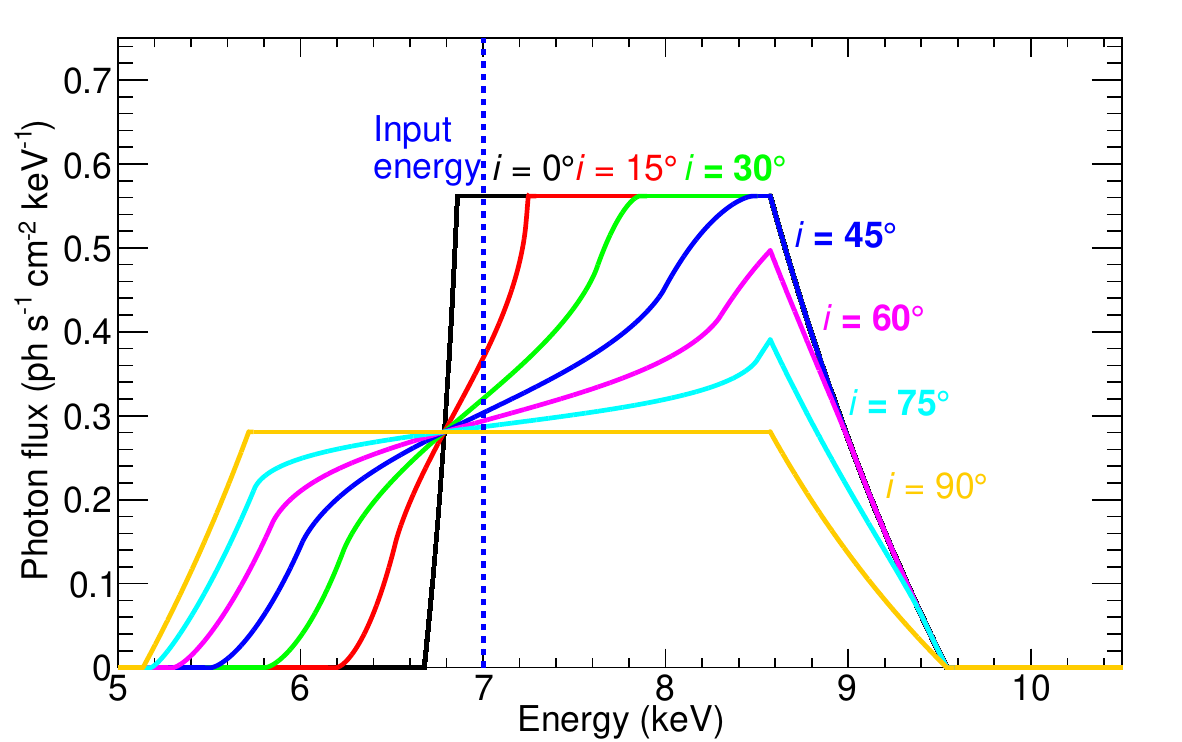}
    \caption{
    {\bf\label{fig:conv} Examples of the convolved emission line profile.}
    {\it Left}: Convolved emission line profiles for an inclination angle of $i=15^{\circ}$. The black lines are the emission line profiles at $E_{0}=7.0{\rm ~keV}$ with a photon flux of $N_{0}=1{\rm ~ph~s^{-1}~cm^{-2}~keV^{-1}}$ convolved with Eq.~\ref{eq:conv} for outflowing velocities ranging from $0.2c$ to $0.3c$. The red line represents the averaged line profile, considering equal contributions from velocities of $0.2c$--0.3$c$.
    {\it Right}: Convolved emission line profiles for different inclination angles. The energy and velocities are the same as in the left figure.
    }    
\end{figure*}

Since the ionization parameters and column densities were tied to the absorption model in the spectral fitting, the only free parameter of this convolved emission model is normalization.
{The normalization of this model $\kappa$ is the same as the original \textsc{XSTAR} emission model, but since only the emission from the hemisphere is considered here, it should be divided by a factor of 2. Thus the definition of normalization can be expressed as:}
{
\begin{equation}
    \kappa \equiv \frac{f_{\rm cov}}{2}\frac{L_{\rm ion}/10^{38}{\rm ~erg~s^{-1}}}{(D/{\rm kpc})^2}
    \simeq (1.1\pm0.4)\times 10^{-4}\times f_{\rm cov},
\end{equation}
}
where we used a de-boosted ionizing luminosity $L_{\rm ion} = (1.6\pm0.5)\times 10^{46} {\rm ~erg~s^{-1}}$ and luminosity distance $D=860{\rm ~Mpc}$ for cosmological parameters of $H_0=73{\rm ~km~s^{-1}~Mpc^{-1}}$, $\Omega_M=0.27$, and $\Omega_\Lambda=0.73$.
Thus, the covering factor {$f_{\rm cov}(\equiv \Omega/2\pi)$}\footnote{$\Omega$ is the wind solid angle that we can observe (i.e., only the near side of the disk).} can be determined from the emission normalization $\kappa$.

\section*{X-ray spectral analysis}

In order to explore the presence line features above the continuum in a wide energy band, we performed a blind search using an absorbed power-law continuum model and a Gaussian line component, free to be positive or negative.
{We explore a line width of 50 eV to search for slightly broadened features.} The line energy is scanned in the range of 4--10\,keV in the source rest frame, with an energy step of 1\,eV. At each step, we fitted the model and recorded the statistical improvement to the overall fit (if present). Such an improvement can be translated using the F-test in statistical confidence levels for the addition of two more parameters. The background spectrum {was incorporated.}

The line search assuming a width of 50\,eV is capable of detecting most of the features above the continuum, thanks to a combination of intrinsically slightly broadened features and adequate signal-to-noise. As clearly shown in Fig. \ref{fig:scan}, the \resolve\ spectrum reveals a P-Cygni profile with {emission and absorption} in the iron K band. At least five separate absorption lines, spanning from 8.6 to 9.5 keV in the rest frame, populate the band,
{resolving spectral features with a broad and smooth profile when observed by instruments with poor or moderate energy resolution into a forest of well-separated narrower absorption lines.}
The same argument applies to the broad emission feature that spans a width of more than a 2 keV, from about 6 keV up to 8 keV. In both cases, multiple components are detected with confidence levels exceeding 3$\sigma$, and surpassing 5$\sigma$ for the two central absorption peaks.

\begin{figure}
    \centering
    \includegraphics[width=1.\linewidth]{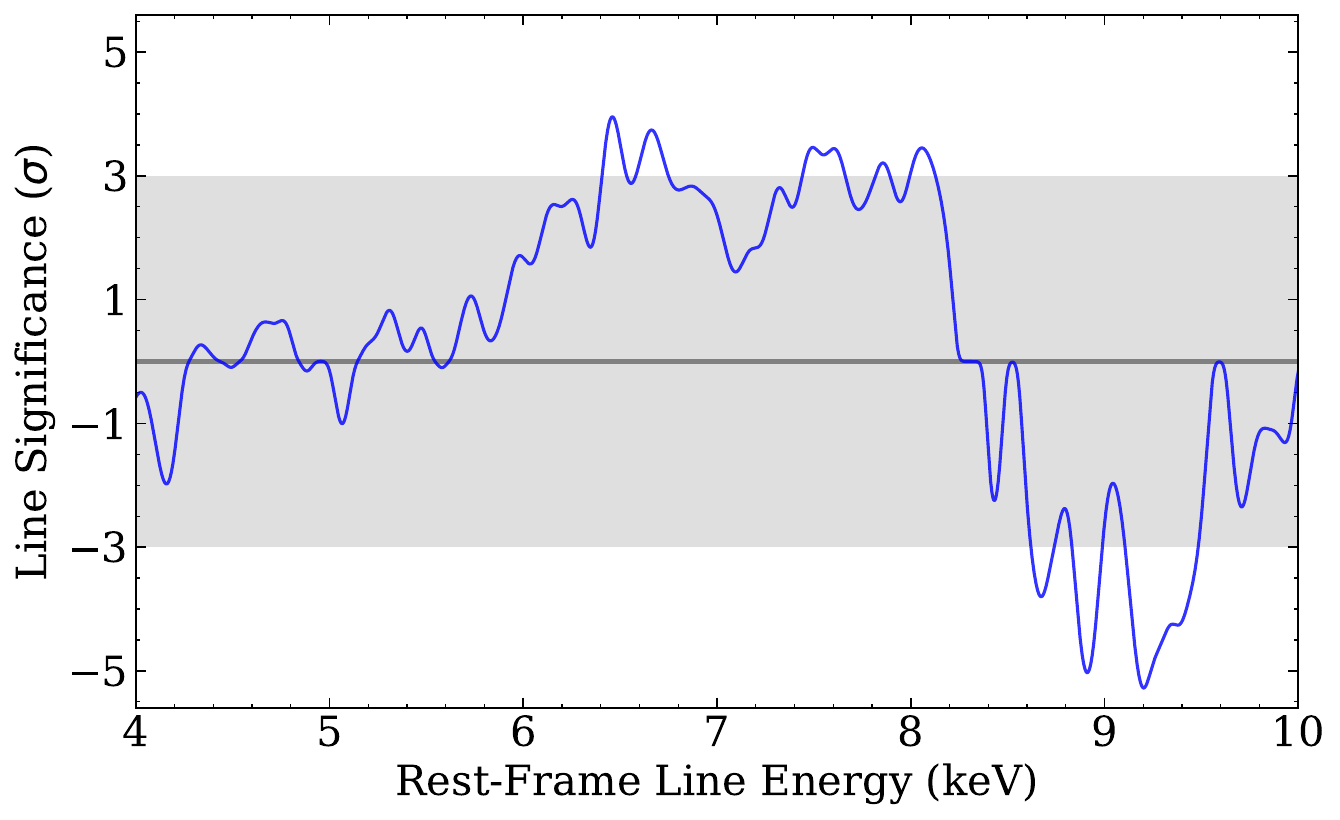}
    \caption{{\bf\label{fig:scan} Result of a blind Gaussian line scan performed on the XRISM \resolve\ observation, plotted in the 4--10 keV range in the PDS 456 rest frame.} {The scan with ${\sigma _{\mathrm{line}} = 50}$ eV is plotted in blue.} The ordinate axis shows the statistical improvement resulting from the addition of the line. Negative values indicate absorption lines. The gray shading represents the 3$\sigma$ confidence level.}
\end{figure}

The average global X-ray spectrum of PDS\,456, covering the 0.5--40\,keV rest frame band, was then fitted, using the mean spectra from \resolve, \xtend\ and \nustar. Data from \xmm\ 
were not included in the time averaged fitting, as these cover only the lower flux interval (iii). Nonetheless, aside from the lower flux during the \xmm\ interval, the overall spectral 
form was consistent with the \xtend\ CCD spectrum.
The \textsc{xstar} absorber models as described above 
should account for the above absorption structure observed in \resolve\ over the 8--10\,keV rest frame band, as well as the overall opacity that is also 
observed in \xtend\ and \nustar\ in the same band. The above emission model was applied in order to account for the broad iron K emission observed in all of the spectra, 
while a lower ionization absorber was also introduced in order to model the pronounced soft X-ray opacity below 4\,keV.  The spectra are fitted using the C-statistic~\cite{Cash1979} and uncertainties 
are given at the 90\% confidence level for 1 interesting parameter (or $\Delta C=2.71$). The \resolve\ NXB model was accounted for in all of the modeling, as is described earlier.

While \resolve\ and \xtend\ covered a strictly simultaneous interval, the \nustar\ spectrum covered the earlier brighter portion of the observations. Given the lack of any spectral 
variability across the observations, this difference was accounted for by introducing a simple multiplicative cross normalisation factor of $\times1.4$ to the \nustar\ data.
The phenomenological form of the model is thus as follows:

\begin{align}
F(E) = & {\rm constant} \times {\rm abs}_{\rm Gal} \times ({\rm abs}_{\rm soft} \times {\rm abs}_{\rm Fe} \times f \times {\rm pow} \nonumber\\
& + [(1-f) \times {\rm pow}] + {\rm emiss}_{\rm soft} + {\rm emiss}_{\rm Fe}).    
\label{eq:model}
\end{align}

\noindent Here the ${\rm constant}$ is the multiplicative factor applied to the \nustar\ data, ${\rm abs}_{\rm Gal}$ is the neutral Galactic absorber (as described in the SED analysis) which absorbs the whole spectrum, while ${\rm abs}_{\rm soft}$ and ${\rm abs}_{\rm Fe}$ are the soft X-ray and iron K-shell 
band outflowing zones of photo-ionized absorption as computed by \textsc{xstar}. These outflowing absorbers then absorb a fraction $f$ of the intrinsic X-ray power-law continuum (${\rm pow}$), while $(1-f)$ represents the fraction of the power-law continuum that is not absorbed. Finally, 
${\rm emiss}_{\rm soft}$ and ${\rm emiss}_{\rm Fe}$ are the velocity broadened 
photo-ionized emission components associated to the soft X-ray and iron K absorbers respectively. As noted above, the column densities and ionization parameters of the emitters were set to be equal to those of the respective absorbers, with the only free parameter being the normalization (flux) of the 
emission. Neither of the emission components are absorbed by the outflowing absorbers. 

\begin{figure}[ht]
\begin{center}
\hspace{-1cm}
\includegraphics[width=\hsize]{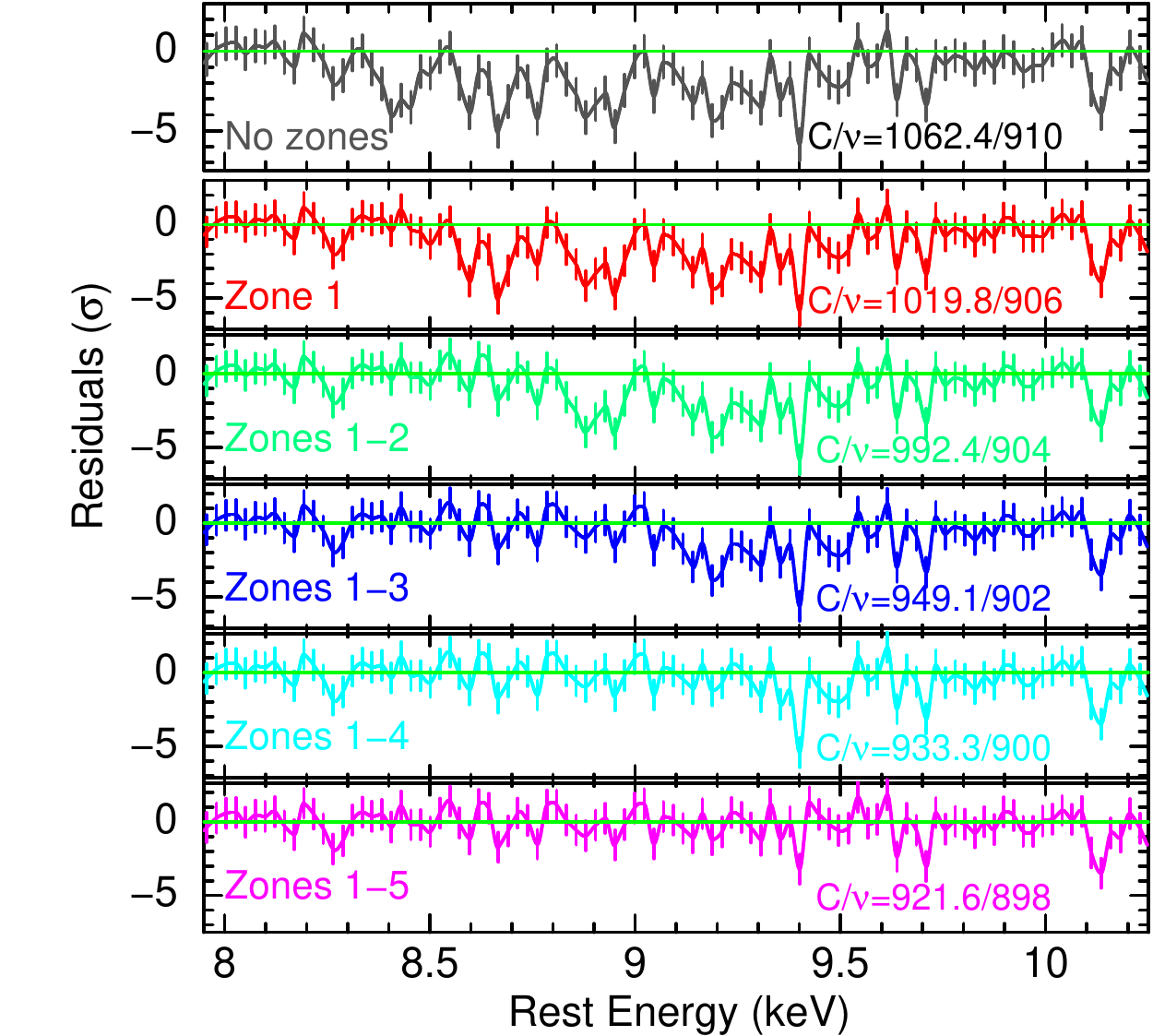}
 \end{center}
\caption{{\bf\label{fig:residuals} Modeling of the \resolve\ spectrum.} The vertical panels show the spectral residuals to the \resolve\ data in the iron K band. This highlights the improvement in the fit statistic as successive outflowing absorber zones (1--5) are each added to the model. The fit statistic at each step is marked.}
\end{figure}

While a single lower ionization absorber is sufficient to model the soft X-ray opacity as measured by \xtend\ towards PDS\,456, successive highly ionized zones of iron K absorption were added 
to the model in order to account for the complex structure observed in the high resolution \resolve\ spectrum over the 8--10\,keV rest frame band (e.g. see Figure~\ref{fig:residuals}). 
Initially all the iron K absorber zones were assumed to be of the same ionization parameter and turbulence, with this value tied between all of these zones, with the column density and outflow velocity allowed 
to vary for each zone. The effect of varying the turbulence and ionization between each zone is investigated later.
The significance of adding each Fe K zone to the model, with respect to just the \resolve\ spectrum, was then calculated by the Akaike Information Criteria\cite{Akaike1974, Tan2012}.
These absorbers, their significances and the parameters are listed in Table~2. 
The fit statistic obtained from fitting the \resolve\ spectrum decreases from $C/\nu=1062.4/910$ to $C/\nu=921.6/898$ upon the addition of the five absorber zones to the model compared 
to the \resolve\ spectrum. For the broad-band 0.5--40\,keV spectrum, the change in fit statistic is even more marked, as this simultaneously accounts for the iron K opacity across all 3 detectors; here the fit statistic decreases from $C/\nu=2234.0/1443$ to $C/\nu=1549.2/1431$ upon the addition of all of the Fe K absorbers. 
For example, even for the weakest zone~5 absorber, the fit statistic improved by $\Delta C=-34.7$ for $\Delta\nu=2$ when measured against the full dataset.  
In this case, the significance of the zone~5 absorber increased from 98\% confidence ({\it Resolve} only) to $>99.99$\% confidence (full dataset). {Indeed, all five zones are detected at $>99.99$\% confidence versus the full dataset; see their significance in parenthesis in Table~2.}

\begin{figure}
\begin{center}
\includegraphics[width=\hsize]{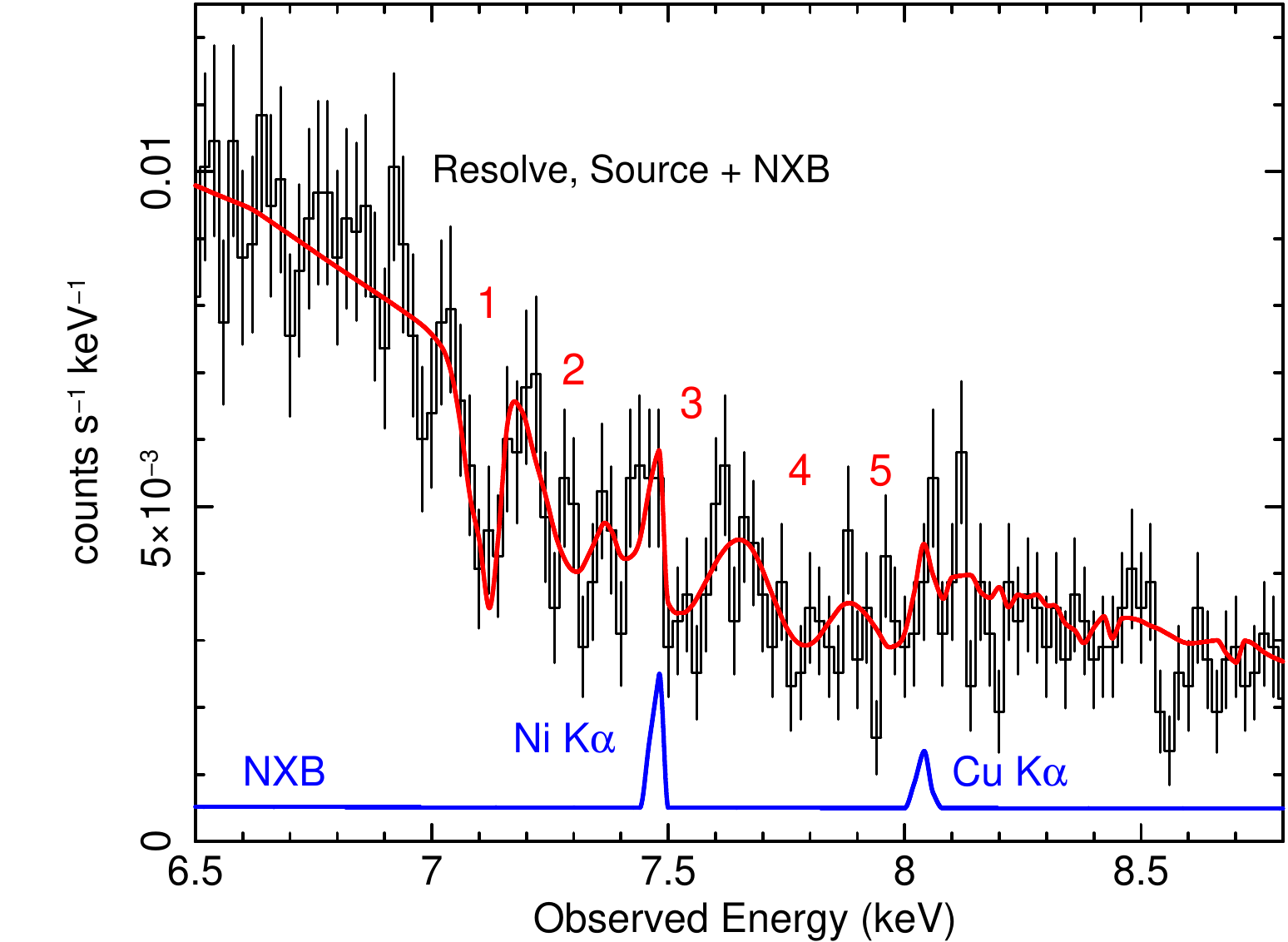}
 \end{center}
\caption{{\bf\label{fig:resolve} The best-fit \textsc{xstar} model applied to the \resolve\ spectrum plotted over the iron K band.}
The total count rate \resolve\ spectrum is shown in black points versus the NXB model (blue), while the five zone \textsc{xstar} model is shown as a red line. Each zone is numerically marked by 
order of increasing velocity. Note that the NXB is predicted to have only a minor impact on the resultant absorption profile, adding a small red-wards peak to absorber zone 3 (from Ni K$\alpha$) and infilling some of absorber on the blue-side of the zone~5 trough (due to Cu K$\alpha$). Note the \resolve\ spectrum is plotted in the observed frame (at $z=0$) for comparison to the NXB model.}
\end{figure}

Figure~\ref{fig:resolve} shows the contribution of each zone to the model, versus that of the \resolve\ NXB, which only has a minimal effect 
on the modeling of the iron K absorption profile. 
The 5 iron-K absorbers cover a wide velocity range from $v/c=-0.226\pm0.002$ to $v/c=-0.333\pm0.001$.
The average value of $v/c=-0.280\pm0.021$, is consistent with previous velocity estimates in PDS\,456 as obtained from a single absorption trough in CCD data \cite{Matzeu2017}.
The best-fit ionization parameter is
found to be $\log\xi=4.90\pm0.14$ when tied across all of the zones, with the dominant opacity arising from He-like iron (Fe\,\textsc{xxv}) for the SED of PDS\,456. 
{If $\xi$ is allowed to vary for each of the five zones, then the fit statistic decreases further to $C/\nu=1529.3/1427$. 
The mean value is then $\log\xi=4.96\pm0.35$, which is consistent with the above value. The larger uncertainty is a result of the variance of $\xi$ across the zones. 
Variations in the turbulence velocity between zones 
are not found to be significant in the \text{xstar} modeling, with $\Delta C<10$ and thus is kept constant between zones.}

The column densities of each iron K zone are similar, with an observed average column density of $N_{\rm H, avg}\approx5\times10^{22}$\,cm$^{-2}$. Note the column density is also subject to special relativistic effects, 
as the outflowing gas sees a smaller proportion of the X-ray continuum as a result of its outwards motion with respect to the source; this is discussed in detail in \cite{Luminari2020}. 
When corrected for this special relativistic aberration, the actual column density is boosted by a factor of $(1 - \beta)/(1 + \beta)$, where $\beta=v/c$ is defined to be negative for outflow towards the observer. The average column density per zone is then $N_{\rm H, corr}\approx1\times10^{23}$\,cm$^{-2}$, with a total column of $N_{\rm H, corr}\approx5\times10^{23}$\,cm$^{-2}$ for all Fe K zones~1--5.

{\color{black}
Figure~\ref{fig:residuals} (lower panel) shows some weak remaining negative residuals in the \resolve\ data compared to the above five Fe K zones model. This is most notable between 8.5--8.6\,keV in the observed frame (or 10.1--10.2\,keV in the rest frame at $z=0.184$). This might suggest the presence of an even higher velocity absorber, at close to $-0.4c$, 
when compared to the expected energy of the resonance absorption due to Fe\,\textsc{xxv} $1s\rightarrow2p$ at 6.7 keV. However the addition of a further, faster absorber only improved 
the fit by $\Delta C=8.0$ for $\Delta\nu=2$ in the \resolve\ data, corresponding to a marginal detection at 95\% confidence. The other remaining residuals are of lower significance. 
}

\begin{figure}[ht]
\begin{center}
\includegraphics[width=\hsize]{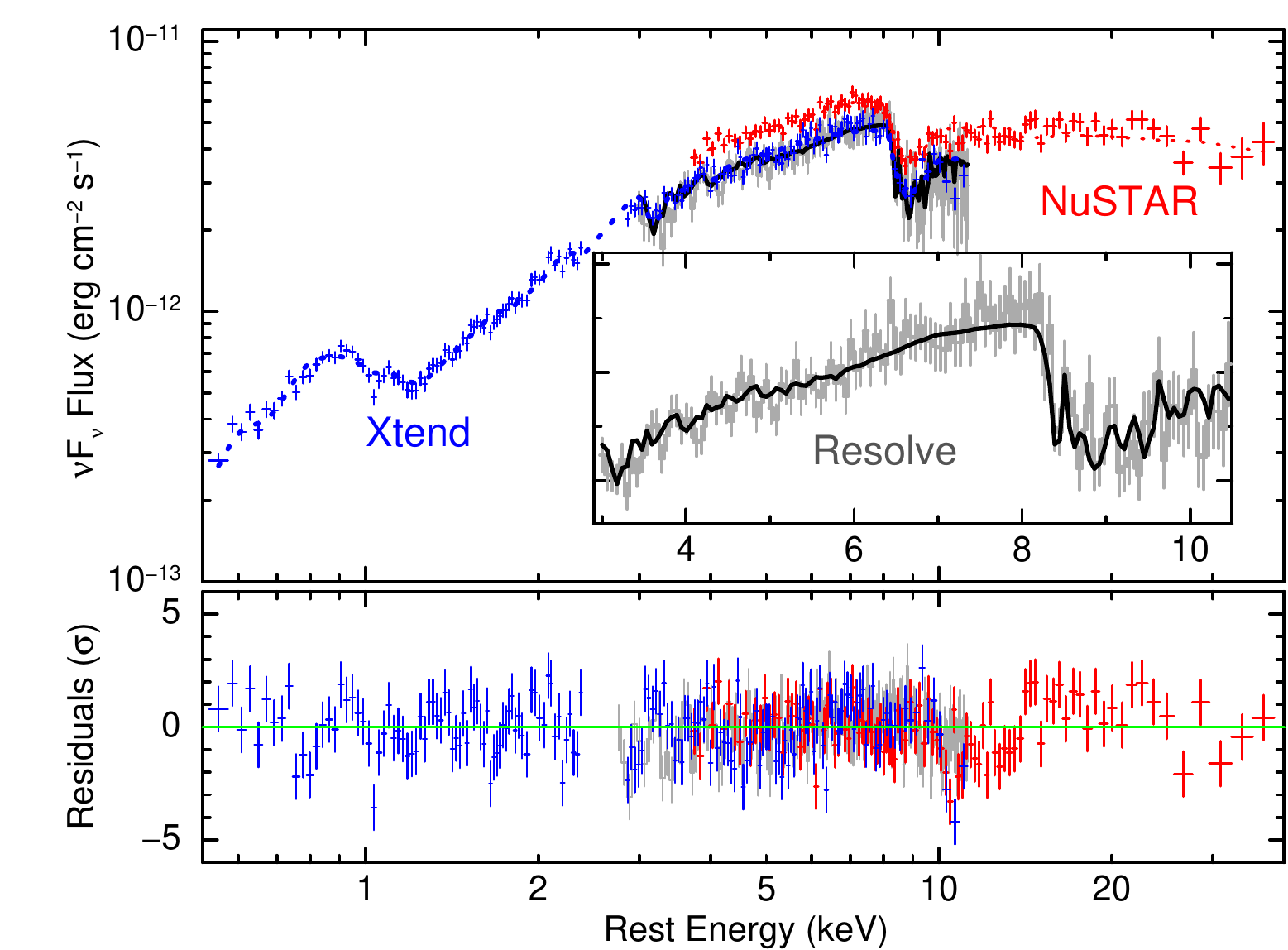}
 \end{center}
\caption{{\bf\label{fig:global} The best-fit model applied to the broad-band spectrum.} The \resolve\ data are shown in black (with grey errors), the \xtend\ data in blue and \nustar\ in red. The upper panel shows the fluxed spectra, folded through the instrumental responses, the lower panel the residuals of the data to the best fitting absorber plus emission model and the inset shows a zoom-in of the model fitted to the 
\resolve\ data over the iron K band.}
\end{figure}

The global model is able to reproduce the overall shape of the broad band X-ray spectrum, as is shown by Figure~\ref{fig:global}. The best-fit photon index, of $\Gamma=2.35\pm0.02$, reproduces the level of the hard X-ray continuum above 10\,keV as observed by \nustar. The soft X-ray absorber, although of similar column to the Fe K absorber zones, is nearly two orders of magnitude lower in its ionization, with $\log(\xi/{\rm erg}\,{\rm cm}\,{\rm s}^{-1})=3.15\pm0.03$. As a result, this absorber is able to reproduce the downwards curvature towards the soft X-ray band as observed in the 
\xtend\ spectrum, which is produced mainly through bound--free opacity in the absorber.  
The line of sight covering fraction of the soft X-ray and iron K absorbers was found to be $f=0.91\pm0.01$, with only 9\% of the power-law continuum remaining un-absorbed. 
The remaining 9\% of continuum flux could either be due to X-ray continuum photons being scattered within the wind, or it might suggest the absorbers partially cover the 
X-ray continuum source. In either case, the absorbing structures must be sufficiently extended in size to absorb the majority of the X-ray coronal emission, which from the X-ray variability is inferred to be 
$10^{15}$\,cm in size (or $16R_{\rm g}$ in gravitational units) and thus provides a geometrical limit on the absorber size-scale. 

Curiously, the outflow velocity of the soft X-ray absorber, of $v/c=-0.277\pm0.001$, is consistent with the average velocity obtained from the iron K absorption zones. 
This suggests it is a lower ionization component associated to the fast wind. 
However, we caution it is not currently possible to determine its velocity from discrete absorption lines in the soft X-rays, as the soft X-ray flux (of $F_{0.4-2.0\,{\rm keV}}=1.0\times10^{-12}$\,erg\,cm$^{-2}$\,s$^{-1}$) is too low to reliably measure these with \xmm\ RGS during interval (iii) of the observations. Potentially, future calorimeter observations at lower energies may be able to resolve multiple component absorption line structures from lighter elements similar to those seen here at iron K.  

The emission from the wind was modeled by a combination of velocity broadened components from the high ionization Fe K emitter (with a total column corresponding to the sum of the zones~1--5) and the lower ionization emitter (with parameters equal to the soft X-ray absorber). The former produces the majority of the broad Fe K emission. 
The geometrical coverage (solid angle) of the emission was calculated from the normalization of the \textsc{xstar} components, as described above. 
From the spectral fits, the normalization of the emitter was found to be 
{$\kappa=(2.1\pm0.3)\times10^{-4}$.} From the above calculation of the emission grids from the SED of PDS\,456, 
this corresponds to a geometrical covering fraction of {$f_{\rm cov}=\Omega/2\pi=1.9\pm0.7$. }
The overall luminosity of the wind emission is $L_{\rm emiss}=(1.5\pm0.3)\times10^{43}$\,erg\,s$^{-1}$ or 2\% of the continuum luminosity.
{
Considering the 90\% uncertainties, the wind coverage is only marginally larger than expected from $2\pi$ coverage of a hemisphere.  
Alternatively, if the normalization is fixed at the expected value for $2\pi$ coverage and the column density of the high ionization emitter is allowed to vary instead, then the column density becomes $N_{\rm H}=(7.5\pm2.5) \times 10^{23}{\rm ~cm^{-2}}$.
This may suggest the total column density over all solid angles is slightly higher than measured along the line of sight. 
Nonetheless, the emission is consistent with a full-covering wide-angle wind, as previously reported in PDS\,456 \cite{Nardini2015}.}

The variability of the broad iron K emission can also give a constraint on its extent and radial location. The spectral intervals (i) through to (iv) were used for this purpose, which following the decline in the continuum following the X-ray flare. Figure~\ref{fig:Feline} shows the variations in the iron K emission, in terms of its flux and equivalent width (EW), with respect to the 2--10\,keV continuum flux. This shows that despite of the factor of $\times4$ decline in the continuum, there was no subsequent change in the emission line flux. Subsequently its equivalent width is anti-correlated with the continuum flux, whereas a constant value would be expected if the emission line flux responded to the continuum variability over the 6 days of the observations. Given the lack of response of the emission line to the continuum, this sets a minimum radial distance scale of $R = c\Delta t \approx 1.5\times10^{16}$\,cm for the emitting gas. 

\begin{figure}
\begin{center}
\includegraphics[width=\hsize]{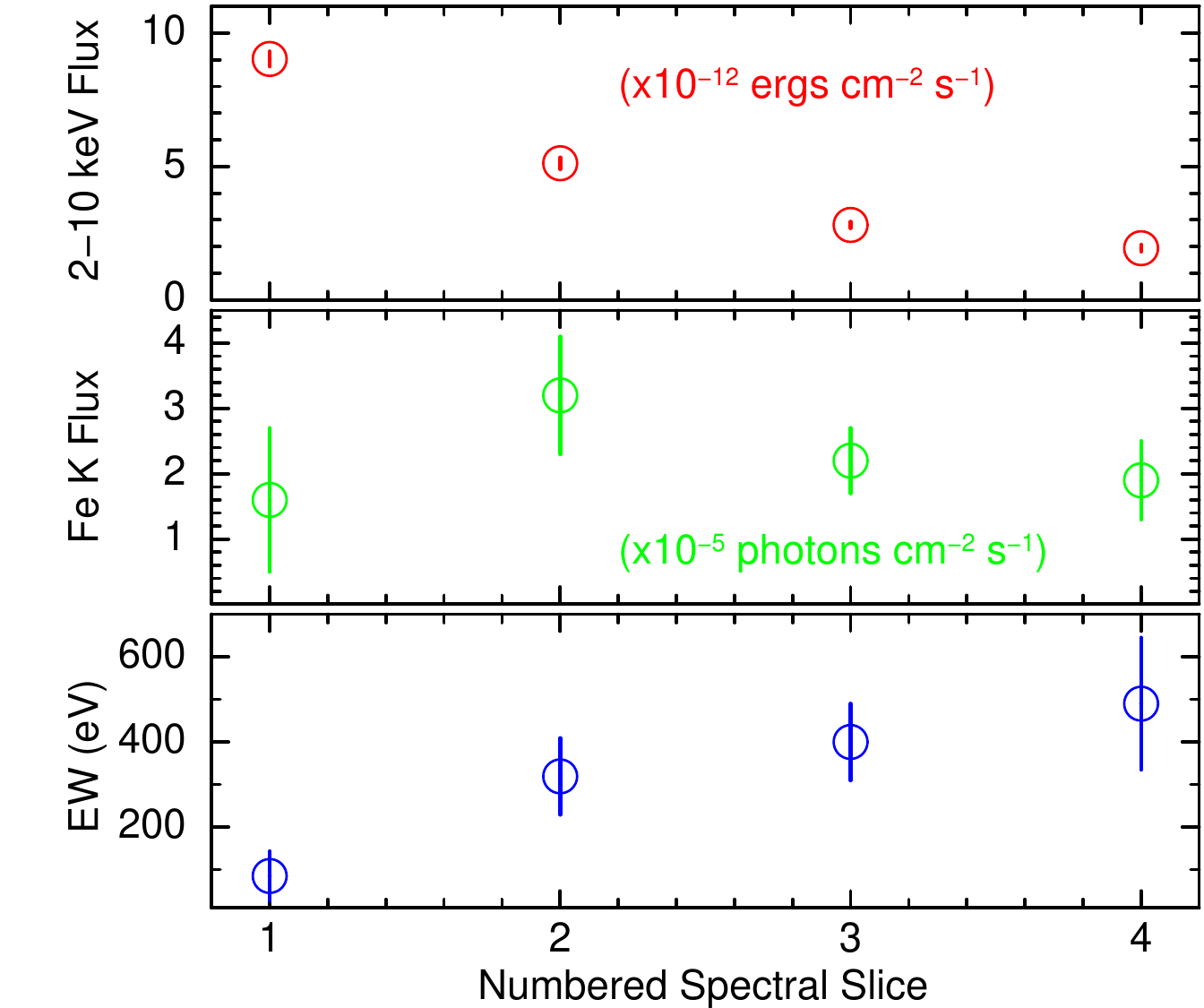}
 \end{center}
\caption{{\bf\label{fig:Feline} The variability of the broad iron K emission in response to the X-ray continuum, plotted for each of the four intervals.}
The lower panel shows the 2--10\,keV power-law flux, versus the iron line flux (middle panel) and 
its equivalent width (EW, lower panel) with respect to the continuum. The flux of the emission line does not appear to respond to the continuum, which results in the EW being anticorrelated with the continuum flux. Thus the size-scale of the emitter is required to be a minimum of 6 light days in extent.}
\end{figure}

Our photoionization absorption results were also independently cross-checked with an alternative photoionization model \textsc{xabs} in \textsc{spex} (v3.08.00~\cite{SPEX}). It adopted the precalculated ionization balance file through the \textsc{spex} task \textsc{xabsinput} with an input SED. Then we applied the code used in Parker~et~al.~\cite{2019Parker} to construct a tabulated \textsc{xabs} model in \textsc{xspec}, including a variety of parameter grids that cover the ranges of those generated by \textsc{xstar}. 

To avoid missing any potential absorbers, we used the codes used in Xu~et~al.~\cite{2023Xu,2024Xu} to perform a systematic grid search over the parameter space upon the baseline model. The baseline model included an absorbed power-law continuum, an emission line component (mentioned above), and a soft X-ray absorber. Therefore, we focused on the search for hard X-ray absorbers. We adopted a logarithmic grid of ionization parameters $\log\xi$ between
1 and 7 with a step of 0.1 and velocities $v_\mathrm{out}$ between $0$ and $-0.4c$ (i.e., only rest frame or outflowing plasma in absorption). The step of $v_\mathrm{out}$ depends on the choice of turbulence velocity $\sigma_\mathrm{turb}$ (300, 700, 1500\,km/s for $\sigma_\mathrm{turb}=100, 1000, 5000$\,km/s). The only free parameter for \textsc{xabs} is the column density, $N_\mathrm{H}$. The $\Delta C\mbox{--}\mathrm{stat}$ was recorded on each grid to reveal the significance of the absorber detection. We found that the results of different $\sigma_\mathrm{turb}$ are consistent, and therefore, the one with $\sigma_\mathrm{turb}=1000$\,km/s is shown in Figure~\ref{fig:xabs-scan}. It clearly reveals five zones for the potential absorption component within the velocity range between -0.22$c$ and -0.33$c$, consistent with those discovered by \textsc{xstar} (see Table~\ref{tab:xstar-fit}). The parameters of 5-zone absorbers derived from \textsc{xabs} are also compatible with the results of \textsc{xstar} within their uncertainties, confirming our spectral fitting results.

\begin{figure}
    \centering
    \includegraphics[width=\hsize]{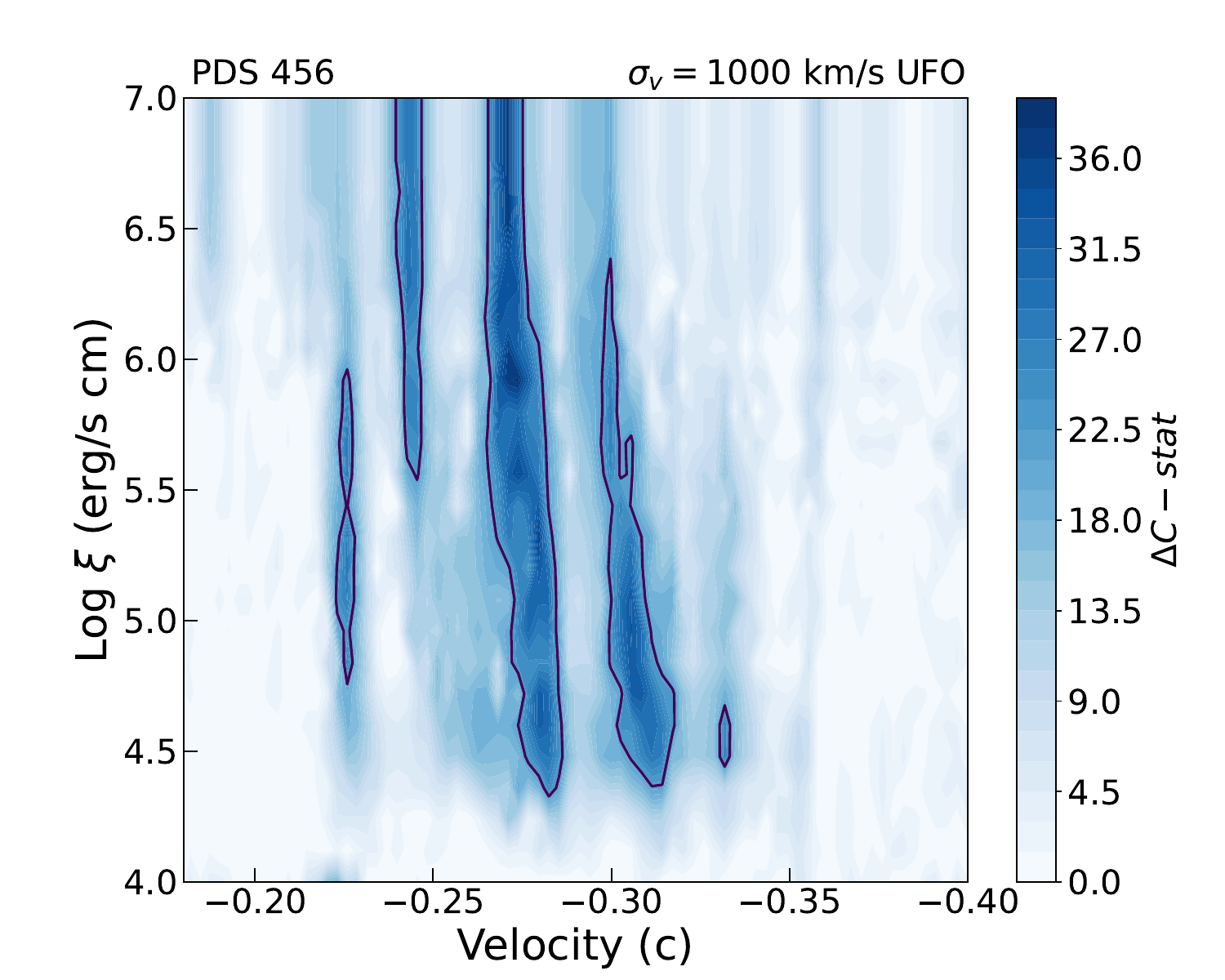}
    \caption{{\bf\label{fig:xabs-scan} Photoionization absorption model scan over the \xrism\ \resolve\ spectra of PDS 456.}
    The color illustrates the statistical improvement after adding an absorption component with a line width of $1000$\,km/s. The \textit{purple} contour corresponds to the $3\sigma$ confidence level. Five potential solutions are revealed in the velocity range between $-0.22c$ and $-0.33c$. }
\end{figure}

A further attempt was pursued to individually constrain the ionization and turbulent velocity of each absorber observed in the \resolve\ data. To do so, we started from the global fitting model outlined above and we ignored \xtend\ and \nustar\ data, since their lower spectral resolution does not allow to spectroscopically resolve the different wind layers. We also replaced the \textsc{XSTAR} Fe K-shell absorbers with \textsc{PHASE} \cite{krongold03} components. Given a tabulated set of ionic abundances as a function of the gas ionization $\xi$ and column density $N_{\rm H}$, PHASE analytically computes the resulting absorption spectrum on-the-fly while performing the spectral fitting as a function of its input parameters, i.e. $\xi, N_{\rm H}, v_{\rm turb}, z$ (the red/blue-shift of the observed features). This allows for a great accuracy when dealing with such high-resolution microcalorimeter data.
The fitting model is the same as Eq.~\ref{eq:model}, but using PHASE components for the Fe K-shell outflowing layers (${\rm abs}_{\rm Fe}$).
Given the limited constraining power of \resolve\ in the soft band, we kept the parameters of the soft wind component, both in absorption and in emission, frozen to the best-fit values of the global fit. We initially left free the Fe K-shell emitter but, since we obtained values consistent with those in the global fit, we fixed them for simplicity. 
To compute the ionic abundances we use the same PDS~456 SED discussed above. The fit was performed with \textsc{XSPEC}. 

As reported in Table \ref{phase_table}, we get the same number of significant components - 5 - and a quite similar range of velocities and total $N_H$. 
However, there is some dispersion between $\log\xi$ and $v_{\rm turb}$ of the individual components and, thus, somewhat different $N_{\rm H}$.
Figure~\ref{phase_fig} shows the best-fit model. It can be seen that each component shows different absorption lines. For layers no. 1, 3, 4, they are Fe~\textsc{xxv} He$\alpha$ and Fe~\textsc{xxvi} Ly$\alpha$: this is due to the higher $\log\xi$ with respect of that in the global fit ($\approx$ 5.5 vs. 4.9), which leads to comparable Fe~\textsc{xxv} and \textsc{xxvi} abundances. For layers no. 2, 5, the most abundant Fe ions are instead \textsc{xxiii}, \textsc{xxiv}, \textsc{xxv}  
(with corresponding columns of 1.3, 2.3, 5.6 $\times 10^{17} {\rm ~cm^{-2}}$, respectively).
We get a fraction of $\sim 10 \%$ of unabsorbed power-law continuum, consistent with the above results.

\begin{figure}
\begin{center}
{\includegraphics[width=1.05\columnwidth]{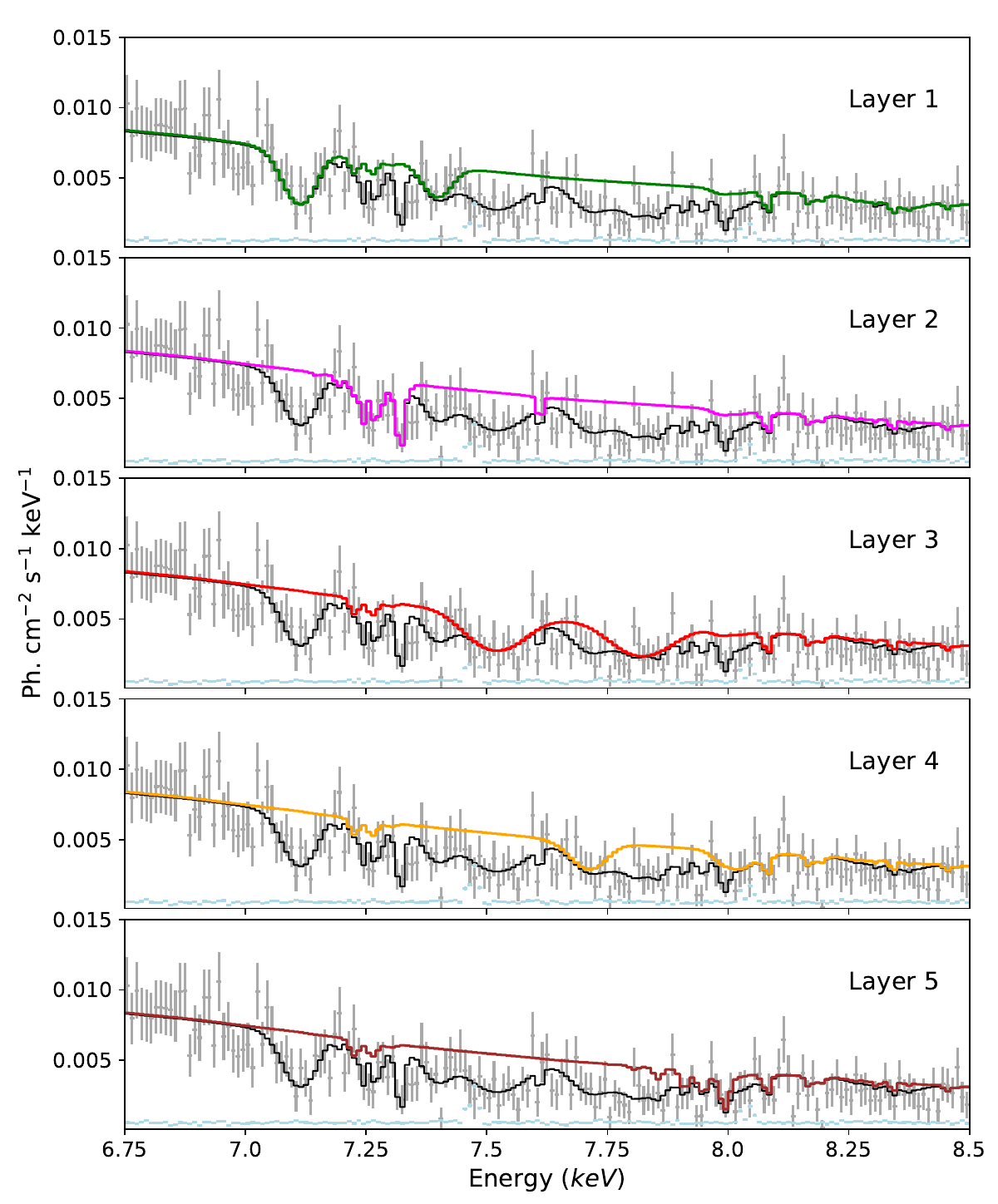}}
 \end{center}
\caption{{\bf\label{phase_fig} The PHASE best-fit model to the \resolve\ data in the 6.75--8.5 keV band.} For each panel, black line reports the best fit, while the colored line outlines the contribution of each component (Layer 1 to 5 from top to bottom). Data and background are reported with grey and lightblue points, respectively.}
\end{figure}

\begin{figure}
\begin{center}
\includegraphics[width=\hsize]{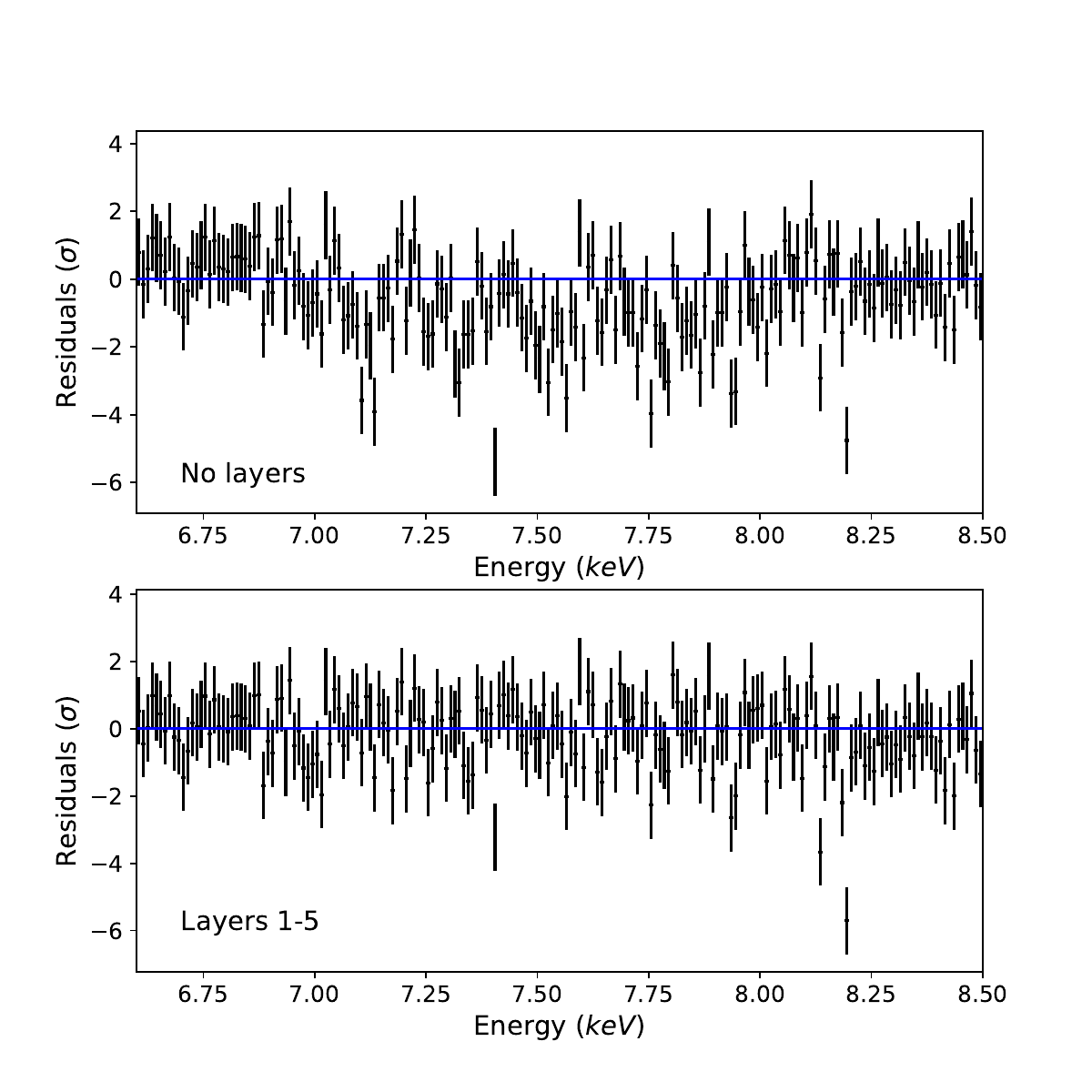}
 \end{center}
\caption{{\bf\label{phase_res_fig} The model-to-data residuals in the 6.7--8.5 keV \resolve\ band when no absorber is included (top) and including all the 5 PHASE layers (bottom).} }
\end{figure}

Table~\ref{tab:param_summary} reports the main parameters of the different fitting models. We listed here only the parameters relevant for the determination of the wind clumpiness and energetic. The differences between models are less than 10\%, indicating the systematic uncertainties due to the choice of model are negligible to those aims. Therefore, we use $\log(\xi/{\rm erg\,s^{-1}\,cm})=5$, $N_{\rm H}=10^{23}{\rm ~cm^{-2}}$ per clump, and $v_{\rm out}=-0.28c$ as fiducial parameters for the discussion.

\end{methods}

\renewcommand{\arraystretch}{0.8}
\begin{table*}
\caption{Summary of PDS~456 Observation Campaign}
\centering
\scriptsize
\begin{tabular}{lcccccc}
\hline\hline
Telescope & Obs. ID & Instrument & Start Time (UT) & End Time (UT) & Duration (ks) & Exposure (ks) \\
\hline
\xrism\ & 300072010 & \resolve\ & 2024-03-11 01:16:50 & 2024-03-17 01:19:08 & 518.5 & 258.1\\
	    & 			 & \xtend\ & 2024-03-11 00:50:02 & 2024-03-17 01:16:55 & 520.0 & 223.9\\
\hline
\nustar\ & 60901011002 & FPMA & 2024-03-10 23:21:09 & 2024-03-14 17:41:09 & 325.2 & 159.2 \\
& & FPMB & 2024-03-10 23:21:09 & 2024-03-14 17:41:09 & 325.2 & 157.3 \\
\hline
\xmm\ & 0931190101 & EPIC-pn/MOS, & 2024-03-13 00:40:50 & 2024-03-14 11:05:03 & 123.9 & 83.1$^a$ \\
 & & RGS, OM & & & & \\
\hline
\swift$^b$ & 00010383121 & XRT, UVOT &  2024-03-01 08:38:39 & 	2024-03-01 09:05:54 & 1.6 & 1.6 \\
	& 00010383122 & XRT, UVOT & 	2024-03-02 08:16:19 & 	2024-03-02 08:43:53 & 1.7 & 1.6 \\
	& 00010383123 & XRT, UVOT & 	2024-03-05 23:20:08 & 	2024-03-05 23:47:52 & 1.7 & 1.7 \\
	& 00010383124 & XRT, UVOT & 	2024-03-06 08:44:03 & 	2024-03-06 08:58:53 & 0.9 & 0.9 \\
	& 00010383124 & XRT, UVOT & 	2024-03-06 19:48:53 & 	2024-03-06 20:02:53 & 0.8 & 0.8 \\
	& 00010383125 & XRT, UVOT & 	2024-03-07 00:52:57 & 	2024-03-07 01:01:54 & 0.5 & 0.5 \\
	& 00010383125 & XRT, UVOT & 	2024-03-07 05:15:19 & 	2024-03-07 05:42:52 & 1.7 & 1.6 \\
	& 00010383126 & XRT, UVOT & 	2024-03-08 03:39:23 & 	2024-03-08 03:45:52 & 0.4 & 0.4 \\
	& 00010383126 & XRT, UVOT & 	2024-03-08 12:52:35 & 	2024-03-08 13:20:53 & 1.7 & 1.7 \\
	& 00010383127 & XRT, UVOT & 	2024-03-09 03:10:50 & 	2024-03-09 03:38:52 & 1.7 & 1.7 \\
	& 00089654001 & XRT, UVOT & 	2024-03-10 13:54:24 & 	2024-03-10 14:21:54 & 1.6 & 1.6 \\
	& 00089654001 & XRT, UVOT & 	2024-03-11 02:55:59 & 	2024-03-11 03:04:53 & 0.5 & 0.5 \\
	& 00010383128 & XRT, UVOT & 	2024-03-11 05:57:49 & 	2024-03-11 06:05:53 & 0.5 & 0.5 \\
	& 00010383128 & XRT, UVOT & 	2024-03-11 15:21:27 & 	2024-03-11 15:45:54 & 1.5 & 1.5 \\
	& 00010383129 & XRT, UVOT & 	2024-03-12 00:55:21 & 	2024-03-12 01:17:50 & 1.3 & 1.3 \\
	& 00010383129 & XRT, UVOT & 	2024-03-12 18:15:22 & 	2024-03-12 18:22:54 & 0.5 & 0.4 \\
	& 00010383130 & XRT, UVOT & 	2024-03-13 08:20:49 & 	2024-03-13 08:46:53 & 1.6 & 1.6 \\
	& 00010383130 & XRT, UVOT & 	2024-03-13 11:44:41 & 	2024-03-13 11:52:53 & 0.5 & 0.5 \\
	& 00010383131 & XRT, UVOT & 	2024-03-14 09:36:44 & 	2024-03-14 10:04:52 & 1.7 & 1.7 \\
	& 00010383132 & XRT, UVOT & 	2024-03-15 03:20:43 & 	2024-03-15 03:26:52 & 0.4 & 0.4 \\
\hline
\textit{NICER} & 7204240101 & XTI & 	2024-03-10 18:54:20 & 	2024-03-10 19:11:40 & 1.0 & 0.8 \\
	& 7204240102 & XTI & 	2024-03-11 13:32:22 & 	2024-03-11 16:59:20 & 12.4 & 1.9 \\
	& 7204240104 & XTI & 	2024-03-14 02:05:12 & 	2024-03-14 20:51:40 & 67.6 & 4.4 \\
	& 7204240105 & XTI & 	2024-03-15 01:30:00 & 	2024-03-15 23:33:20 & 79.4 & 5.2 \\
	& 7204240106 & XTI & 	2024-03-16 00:45:40 & 	2024-03-16 14:43:00 & 50.2 & 3.8 \\
\hline
\textit{Seimei}  & --- & KOOLS-IFU & 2024-03-09 19:04:44 & 2024-03-09 20:22:35 & 4.7 & 4.7\\
	     & --- & KOOLS-IFU & 2024-03-10 19:40:27 & 2024-03-10 20:23:15 & 2.6 & 2.6 \\
	     & --- & TriCCS         & 2024-03-13 19:33:04 & 2024-03-13 19:43:04 & 0.6 & 0.6 \\
	     & --- & KOOLS-IFU & 2024-03-14 19:52:07 & 2024-03-14 19:57:26 & 0.3 & 0.3 \\
	     & --- & KOOLS-IFU & 2024-03-15 20:05:43 & 2024-03-15 20:11:02 & 0.3 & 0.3 \\
\hline\hline
\end{tabular}
\begin{flushleft}
\footnotesize
$^a$ Exposure time for EPIC-pn.\\
$^b$ Only observations in March 2024 are listed, while the entire monitoring campaign includes more observations before and after the \xrism\ observation.
\end{flushleft}
\label{tab:obs}
\end{table*}

\begin{table*}
\caption{\label{tab:xstar-fit}Photoionization modeling with \textsc{xstar} for the PDS\,456 wind components. }
\centering
\scriptsize
\begin{tabular}{lccccccc}
 \hline\hline
ZONE &$N_{\rm H,obs}$ &$v_{\rm {out}}/c$  &log$\xi$$^a$    &$\sigma_{\rm {turb}}$$^b$ & $N_{\rm H,corr}$$^c$ &$\Delta C/\Delta\nu$$^{d} $  &$P_{\rm{null}}^e$\\
 &($\times 10^{22}$ cm$^{-2}$)  &&&(km s$^{-1}$) & $(\times 10^{22}$ cm$^{-2}$) &&\\
 \hline\hline
 1  &   5.0$\errUD{1.4}{1.2}$ &  $-0.226\errUD{0.002}{0.002}$   & $4.90\errUD{0.14}{0.14}$   & $1900\errUD{600}{400}$    &  7.9$\errUD{2.2}{1.9}$   & $-42.6/4$ ($-378.7/4$) & $3.1\times 10^{-8}$ ($3.2\times 10^{-81}$)\\
 2 &  4.9$\errUD{1.4}{1.4}$  &   $-0.254\errUD{0.002}{0.002}$ &  $4.90^t$    &  $1900^t$     & 8.2$\errUD{2.4}{2.4}$  & $-27.4/2$ ($-92.3/2$) & $8.3\times 10^{-6}$ ($6.7\times 10^{-20}$)  \\
 3          &  7.6$\errUD{2.0}{1.5}$     & $-0.278\errUD{0.001}{0.001}$ & $4.90^t$  &   $1900^t$    &  13.5$\errUD{3.5}{2.7}$    & $-43.3/2$ ($-128/2$) & $2.9\times 10^{-9}$ ($1.2\times 10^{-27}$)   \\
 4 &  4.5$\errUD{1.4}{1.3}$    &$-0.307\errUD{0.003}{0.003}$ &  $4.90^t$      &   $1900^t$  &   8.5$\errUD{2.6}{2.5}$  &  $-15.8/2$ ($-51.6/2$) & $2.7\times 10^{-3}$ ($4.6\times 10^{-11}$)  \\
5&  4.9$\errUD{1.5}{1.2}$ & $-0.333\errUD{0.001}{0.001}$  &$4.90^t$  & $1900^t$                &  9.8$\errUD{3.0}{2.4}$  &  $-11.7/2$ ($-34.7/2$) & $2.1\times 10^{-2}$ ($2.2\times 10^{-7}$)  \\
 \hline\\ 
TOTAL & 26.9$\errUD{3.5}{3.0}$ &--&--&--&47.9$\errUD{6.2}{5.3}$  &--&--\\
MEAN & 5.4$\errUD{0.7}{0.6}$ &$-0.280\errUD{0.021}{0.021}$$^f$ &-- &--&$9.6\errUD{1.2}{1.1}$  &--&--\\
  \hline\\ 
SOFT &10.5$\errUD{0.5}{0.5}$ &$-0.277\errUD{0.001}{0.001}$& $3.15\errUD{0.03}{0.03}$ &$<600$ & $18.5\errUD{0.9}{0.9}$  &--&--\\
\hline  \hline\\
\end{tabular}
\begin{flushleft}
\footnotesize
$^a$ Ionization parameter in units of erg\,cm\,s$^{-1}$. Values are tied to zone\,1. Note if $\log\xi$ is allowed to vary independently 
for zones 1--5, then the mean value is $\log\xi=4.88\pm0.24$, consistent with the tabulated value.\\
$^b$ The absorber turbulence velocity or line width in terms of the $1\sigma$ Gaussian width.\\ 
$^c$ The column density, $N_{\rm H,corr}$, corrected for relativistic effects; see \cite{Luminari2020}.\\
$^d$ The $\Delta C$ statistic is reported with respect to the \resolve, $\Delta\nu$ is the change in the degrees of freedom upon the addition of each zone to the model. {Values in parenthesis are with respect to the global, broad-band fit.}\\
$^e$ Null probability for the significance of successively adding each zone, 1 through to 5, to the model against the {\it Resolve} data {(values in parenthesis are against the global fit)}. 
This is calculated using the AIC method, whereby $P = e^{\Delta {\rm AIC}/2}$ and $\Delta {\rm AIC} = \Delta C + 2\Delta\nu$.\\
$^f$ Mean velocity errors were calculated from the variance of individual values (i.e. standard error), rather than from the propagation of individual errors.\\
$^t$ Denotes that the parameter was tied.
\end{flushleft}
\end{table*}

\begin{table*}
\caption{\label{phase_table}Best-fit values of the PHASE fit to the \resolve\ data in the 2.2--9.5 keV band. Errors are reported at 90\% c.l.}
\scriptsize
\begin{tabular}{c c c c c c c }
\hline\hline
Power-law & $\Gamma$ & norm (abs.)$^a$ & norm (unabs.)$^a$& & \\
 & $2.3 \pm 0.1$ & $(3.4 \pm 0.8) \times 10^{-3}$ & $(2.6^{+ 2.3}_{-1.3}) \times 10^{-4}$ & \\
 \hline\hline
PHASE & $\log\xi$$^b$ & $\log N_{\rm H,obs}$$^c$ & $\log N_{\rm H,corr}$$^c$ & $v_{\rm out}$ ($c$) & $v_{\rm turb}$ (km s$^{-1}$) & $\Delta C / \Delta \nu$$^e$ \\
\hline
Layer 1 & $5.5^{+0.3}_{-0.5}$ & $22.8 \pm 0.2$ & $23.0 \pm 0.2$ & 0.226 $\pm 0.001$ & $1489^{+877}_{-745}$ & 35.8/4  \\
Layer 2 & $4.6 \pm 0.2$ & $22.5^{+0.2}_{-0.3}$ & $22.7^{+0.2}_{-0.3}$ & 0.253 $\pm 0.001$ & $<61$ & 20.5/4 \\
Layer 3 & $5.8 \pm 0.2$ & $23.2 \pm 0.2$ & $23.4 \pm 0.2$ & 0.278 $\pm 0.003$ & $2889^{+1696}_{-789}$ & 35.0/4 \\
Layer 4 & $5.5 \pm 0.4$ & $22.6^{+0.3}_{-0.4}$ & $22.9^{+0.3}_{-0.4}$ & 0.301 $\pm 0.002$ & $1632^{+1933}_{-987}$ & 39.0/4 \\
Layer 5 & $4.5^{+0.4}_{-0.2}$ & $22.3^{+0.2}_{-0.4}$ & $22.6^{+0.2}_{-0.4}$ & 0.333 $\pm 0.001$ & $<1376$ & 37.2/4 \\
Total & & 23.5 & 23.7 & & & 167.5/20 \\
\hline\hline
\end{tabular}
\begin{flushleft}
\footnotesize
$^a$ normalisation in units of photons keV$^{-1}$ cm$^{-2}$ s$^{-1}$ at 1 keV.\\
$^b$ Ionization parameter in units of ${\rm erg~cm~s^{-1}}$.
$^c$ Observed column density in units of ${\rm  cm^{-2}}$.\\
$^d$ Intrinsic column density in units of ${\rm  cm^{-2}}$ (i.e., corrected for the relativistic effects).\\
$^e$ The change in the fit statistics and degrees of freedom upon the addition of each zone to the model.
\end{flushleft}
\end{table*}

\begin{table*}
\caption{\label{tab:param_summary} Summary of parameters among different models.}
\centering
\scriptsize
\begin{tabular}{lccccc}
 \hline\hline
Models & Number of zones & Mean log$\xi$$^a$ & Total $N_{\rm H~corr}$$^b$ & Mean $N_{\rm H~corr}$$^b$ & Mean $v_{\rm {out}}$\\
& & & ($\times 10^{22}$ cm$^{-2}$) & ($\times 10^{22}$ cm$^{-2}$) & ($c$) \\
 \hline
\textsc{XSTAR}/\textsc{XABS} & 5 & $4.90\errUD{0.14}{0.14}$ & 47.9$\errUD{6.2}{5.3}$ & $9.6\errUD{1.2}{1.1}$ &$-0.280\errUD{0.021}{0.021}$ \\
\textsc{PHASE} & 5 & $5.18\errUD{0.14}{0.15}$ & $52.1\errUD{1.8}{1.1}$ & $10.4\errUD{3.6}{2.3}$ &$-0.278\errUD{0.019}{0.019}$ \\
\hline  \hline\\
\end{tabular}
\begin{flushleft}
\footnotesize
$^a$ Ionization parameter in units of erg\,cm\,s$^{-1}$.\\
$^b$ The column density, $N_{\rm H}$, corrected for relativistic effects.\\
\end{flushleft}
\end{table*}
\end{document}